\documentclass[aip,reprint]{revtex4-1}
\usepackage{color} 
\usepackage{amssymb,amsmath} 
\usepackage{graphicx}
\usepackage{subfigure}
\usepackage{enumerate}
\usepackage{ulem}

\begin{document}

\title{Electronic excitation dynamics in multichromophoric systems described via a polaron-representation master equation}
\author{Avinash Kolli}
\affiliation{Department of Physics and Astronomy, University College London, Gower Street, London WC1E 6BT, United Kingdom}
\author{Ahsan Nazir}
\affiliation{Department of Physics and Astronomy, University College London, Gower Street, London WC1E 6BT, United Kingdom}
\affiliation{Blackett Laboratory, Imperial College London, London SW7 2AZ, United Kingdom}
\author{Alexandra Olaya-Castro}\email{a.olaya@ucl.ac.uk}
\affiliation{Department of Physics and Astronomy, University College London, Gower Street, London WC1E 6BT, United Kingdom}

\definecolor{purple}{RGB}{102,0,153}

\begin{abstract}

We derive a many-site version of the non-Markovian  time-convolutionless  polaron master equation [S. Jang et al., J. Chem Phys. \textbf{129}, 101104 (2008)] to describe electronic excitation dynamics in multichromophoric systems. By treating electronic and vibrational degrees of freedom in a combined frame (polaron frame), this theory is capable of interpolating between weak and strong exciton-phonon coupling and is able to account for initial non-equilibrium bath states and spatially correlated environments. Besides outlining a general expression for the expected value of any electronic system observable in the original frame, we also discuss implications of the Markovian and Secular approximations highlighting that they need not hold in the untransformed frame despite being strictly satisfied in the polaron frame. The key features of the theory are illustrated using as an example a four-site subsystem of the Fenna-Mathews-Olson light-harvesting complex. For a spectral density including a localised mode, we show that oscillations of site populations may only be observed when non-equilibrium bath effects are taken into account. Furthermore, we illustrate how this formalism allows us to identify the electronic and vibrational components of the oscillatory dynamics. 

\end{abstract}

\maketitle

\section{Introduction}

Electronic resonance energy transfer is a widespread phenomenon in a variety of systems ranging from biomolecular components of the photosynthetic machinery \cite{sundstrom97, renger01,vangrondelle06, cheng09}, DNA \cite{dna} and fluorescence-based sensors \cite{sensors} to conjugate polymers \cite{polymers1, polymers2}, crystal impurities\cite{soules71,rackovsky73} and quantum dot arrays \cite{crooker02,gerardot05,kim08}. Traditionally, electronic energy transfer  in many of these systems has been described with F\"orster-Dexter theory \cite{foerster59,dexter52} which gives account of a Pauli-type dynamics of the probabilities of individual chromophores being excited or de-excited, as if an excitation were ``hopping'' around. This theory has been widely applied \cite{scholes03,beljonne09} and is particularly successful in describing energy transfer when the electronic coupling between chromophores is very weak in comparison to their interaction with fast-relaxing vibrational degrees of freedom. Outside this limit, the electronic interaction becomes significant in comparison to exciton-phonon coupling and quantum effects can take place, producing for example, collective (excitonic) behaviour \cite{scholes01}, coherent exciton dynamics \cite{olayacastro08, mulken07} or interference of energy transfer pathways \cite{hoda11}. Therefore, it is of considerable interest to have a theoretical framework encompassing in a unified manner the various types of electronic excitation dynamics that can be observed in multichormophoric assemblies. 

The interest in such a unified framework, and particularly in coherence effects, has been further motivated by recent experimental works probing incisively ultrafast excitation dynamics and witnessing coherent evolutions of excitonic superpositions in light-harvesting antennae \cite{engel07,collini09,mercer09,collini10,engel10}. A common approach to study the effects of coherence in exciton relaxation has been the use of Redfield type master equations \cite{breuer07,may04}, where the exciton-phonon  coupling is the smallest energy scale in the open quantum system and therefore can be treated as a perturbation to derive a master equation for the electronic degrees of freedom. This approach has been used to provide insights into the effects of coherence in spectroscopic signals \cite{kuhn02, may04} and more recently to understand the roles of coherence and relaxation in the efficiency of excitation transfer \cite{mohseni08,rebentrost09a,rebentrost09b,caruso09a,olayacastro09,sarovar09,rebentrost09c,caruso10,olayacastro10}.  However, many multichromophoric systems operate in an intermediate regime where exciton, vibronic relaxation and exciton-phonon coupling energy scales are comparable and hence traditional perturbative treatments become inaccurate. This has lead to recent investigation of excitation dynamics using non-perturbative approaches \cite{ishizaki09a,ishizaki09b,irene09,prior10,thorwart09,nalbach10,nalbach11} and sophisticated stochastic treatments \cite{eisfield09} of the system-plus-bath dynamics.  Though accurate for small aggregates, non-perturbative calculations become very inefficient as the system size increases, or in the case of multiple-excitations even if the number of chropmophores is small.  It is therefore of much relevance to develop modified perturbative methodologies \cite{jang08, jang09, nazir09, mccutcheon11}  that can provide an appropriate qualitative and quantitative account of dynamics in the intermediate regime whilst being computationally tractable. 

In this work we generalize to multichromophoric aggregates the polaron-modified master equation formalism pioneered by Abram and Silbey \cite{silbey75, silbey89} and studied recently by Jang et al. \cite{jang08} and Nazir \cite{nazir09}. Although perturbative, this approach interpolates between the two limits of weak and strong exciton-phonon coupling, allowing for a consistent exploration of the regime where the energy scales of electronic coupling and exciton-bath interaction are comparable.  In this formalism, the electronic system-plus-phonon bath Hamiltonian is transformed into a new frame (polaron frame) where electronic couplings are renormalized and fluctuate due to the interaction with the vibrational modes, which are in turn fully displaced due to the interaction with the electronic excitation. In this way not only is the effect of the bath on the electronic system considered, but also a reciprocal effect on the phonon bath is accounted for.  Under certain conditions, the energy scale of the electronic coupling fluctuations induced by the displaced vibrations is small in comparison to all other energy scales in the system and, therefore, such fluctuations can be treated as a perturbation. Standard projection operator techniques can then be used to derive a second-order master equation that captures non-Markovian and non-equilibrium bath effects in the intermediate regime. In addition to the full derivation of the homogenous and inhomogeneous terms of the time-convolutionless polaron master equation for single-exciation dynamics in a multichromophoric aggregate, this paper contributes the following results:

\begin{enumerate}[(i)]
\item We present a general framework for evaluating all electronic system observables in the original untransformed lab frame. This extends previous work \cite{jang08,jang09}, which have only been able to consider the dynamics of site populations. Full knowledge of the reduced electronic density operator in the original frame will allow comparison of  the non-Markovian theory outlined here with experimental data. 
\item We illustrate the versatility of the theory to capture the dynamical effects induced by structured harmonic environments. Particularly, we use as an example a four-site subsystem of the Fenna-Matthews-Olsen (FMO) complex and show that non-equilibrium bath effects, captured by the inhomogeneous term, are crucial for the observation of long-lived site population oscillations in the presence of localised vibrational modes. In general, the theory can capture the interplay between electronic interactions and localised vibrations in excitation dynamics, allowing us to elucidate the possible electronic and vibronic origin of the oscillations.  
\item We apply two common approximations to the system dynamics in the polaron frame: the Born-Markov approximation and the secular approximation. Importantly, we are able to show that approximations made within the polaron frame do not necessarily hold within the original untransformed lab frame; hence, Markovian dynamics in the polaron frame may capture some non-Markovian effects in the lab frame. Finally, we show that the theory reduces to F\"orster and Redfield dynamics in the limits of very weak electronic coupling and very weak exciton-phonon coupling respectively.
\end{enumerate}

The paper is organised as follows. In section \ref{sec:2} we begin by introducing the  multichromophore excitation Hamiltonian and then  derive a non-Markovian master equation describing excitation dynamics within the polaron frame. This section is concluded with a general framework for calculating any expected value of the electronic system in the untransformed lab frame. In section \ref{sec:3} we apply the theory to study the dynamics in a four-site subsystem of the FMO complex that interacts with both continuous and localised vibrational modes. Section \ref{sec:4} discusses implications of the Born-Markov and secular approximations in the polaron frame, as well as the F\"orster and Redfield limits of the theory. Finally, in section \ref{sec:5} we present some concluding remarks. 

\section{Many-Site Polaron Master Equation}
\label{sec:2}

\subsection{Polaron-transformed exciton-phonon Hamiltonian}

In this section, we derive a general master equation describing single-excitation dynamics of $m$ coupled (chromophoric) sites interacting with a common bath of  harmonic oscillators representing the environment (e.g protein and solvent). In order to go beyond the weak system-environment coupling limit, we perform a polaron transformation of the exciton-bath Hamiltonian prior to a perturbative expansion with respect to a re-defined system-environment interaction in the transformed frame~\cite{silbey75,silbey89,jang08,jang09,nazir09,mccutcheon11}. The Born-Markov approximation is avoided and hence both non-Markovian and non-equilibrium environmental effects are accounted for~\cite{jang08,jang09}.

The Hamiltonian describing the combined electronic excitation and harmonic environment in an $m$-site system is ($\hbar=1$):
\begin{eqnarray}
H &=& \sum_{m} \epsilon_{m} \sigma_{m}^{+} \sigma_{m}^{-} + \sum_{\langle m, n \rangle} V_{mn}   (\sigma_{m}^{+} \sigma_{n}^{-} + \sigma_{n}^{+} \sigma_{m}^{-}) \nonumber\\ 
&&\:{+}\sum_{\mathbf{k}} \omega_{\mathbf{k}} b_{\mathbf{k}}^{\dagger} b_{\mathbf{k}}  + \sum_{m} \sigma_{m}^{+} \sigma_{m}^{-} \sum_{\mathbf{k}} (g_{\mathbf{k},m} b_{\mathbf{k}}^{\dagger} + g_{\mathbf{k},m}^{*} b_{\mathbf{k}}).\nonumber\\
\end{eqnarray}
Here $\sigma_{m}^{+}=|m\rangle\langle 0|$ corresponds to creation of an excitation on site $m$ with energy $\epsilon_{m}$, 
$V_{mn}$ denotes the electronic coupling between sites $m$ and $n$, and the notation $\langle m,n \rangle$ signifies a summation over all $m$ and all $n>m$. The operator $b_{\mathbf{k}}^{\dagger}$ ($b_{\mathbf{k}}$) corresponds to the creation (annihilation) operator of the $\mathbf{k}$'th mode of the phonon bath, with frequency $\omega_{\mathbf{k}}$. Finally, $g_{\mathbf{k},m}$ represents the site-dependent coupling of site $m$ to the bath mode $\mathbf{k}$ with associated spectral density $J_m(\omega)=\sum_{\mathbf{k}} |g_{\mathbf{k},m}|^2\delta(\omega-\omega_{\mathbf{k}})$. 

We begin our analysis by moving into the polaron frame defined by the transformation $\tilde{H} = e^{S} H e^{-S}$, where $S=\sum_{m} \sigma_{m}^{+} \sigma_{m}^{-} \sum_{\mathbf{k}}  (\alpha_{\mathbf{k},m} b_{\mathbf{k}}^{\dagger} - \alpha_{\mathbf{k},m}^{*} b_{\mathbf{k}})$, with $\alpha_{\mathbf{k},m} = g_{\mathbf{k},m}/\omega_{\mathbf{k}}$. Within this transformed frame, the Hamiltonian for the single-excitation subspace becomes 
\begin{eqnarray}
\tilde{H}&{}={}&\sum_{m} \tilde{\epsilon}_{m} \sigma_{m}^{+} \sigma_{m}^{-} + \sum_{\mathbf{k}} \omega_{\mathbf{k}} b_{\mathbf{k}}^{\dagger} b_{\mathbf{k}}\nonumber\\  
&&\:{+} \sum_{\langle m, n \rangle} V_{mn}  (B_{mn} \sigma_{m}^{+} \sigma_{n}^{-} + B_{mn}^{\dagger} \sigma_{n}^{+} \sigma_{m}^{-}),
\end{eqnarray}
where the energy of each site is now shifted by its corresponding site-dependent reorganisation energy, $\lambda_{m}=\sum_{\mathbf{k}} \frac{|g_{\mathbf{k},m}|^{2}}{\omega_{\mathbf{k}}}$, such that $\tilde{\epsilon}_{m} = \epsilon_{m} - \lambda_{m}$. Here we have also introduced the new bath operators 
\begin{equation}
B_{mn} = e^{\sum_{\mathbf{k}}(\delta \alpha_{\mathbf{k},m n} b_{\mathbf{k}}^{\dagger} - \delta \alpha_{\mathbf{k},m n}^{*} b_{\mathbf{k}})},
\end{equation}
where $\delta \alpha_{\mathbf{k},m n} = \alpha_{\mathbf{k},m} -\alpha_{\mathbf{k},n}$ depends on the difference in bath couplings of sites $m$ and $n$. We now  separate 
$\tilde{H}$ into two parts, a non-interacting system and bath Hamiltonian
\begin{eqnarray}
\tilde{H}_{0} &=& \sum_{m} \tilde{\epsilon}_{m} \sigma_{m}^{+} \sigma_{m}^{-} + \sum_{\langle m, n \rangle} V_{mn}  \beta_{mn} (\sigma_{m}^{+} \sigma_{n}^{-} + \sigma_{n}^{+} \sigma_{m}^{-}) \nonumber\\ 
&&\:{+}\sum_{\mathbf{k}} \omega_{\mathbf{k}} b_{\mathbf{k}}^{\dagger} b_{\mathbf{k}},
\label{eq:h0}
\end{eqnarray}
and a system-bath interaction described by 
\begin{equation}
\tilde{H}_{I}=\sum_{\langle m, n \rangle} V_{mn}  (\tilde{B}_{mn} \sigma_{m}^{+} \sigma_{n}^{-} + \tilde{B}_{mn}^{\dagger} \sigma_{n}^{+} \sigma_{m}^{-}).
\label{eq:polaronH}
\end{equation}
In doing so, we have defined bath-induced renormalisation factors $\beta_{mn} = \langle B_{mn} \rangle$, and shifted bath operators $\tilde{B}_{mn} = B_{mn} - \beta_{mn}$. This splitting ensures both that (for $\beta_{mn}\neq0$) the coherent transfer dynamics generated by the {\it bath-renormalized} electronic couplings, $\tilde{V}_{nm}=V_{nm}\beta_{nm}$, is fully accounted for within the system Hamiltonian, and that $\langle \tilde{H}_I\rangle=0$. For a harmonic oscillator bath in thermal equilibrium, the renormalisation factors evaluate to $\beta_{mn} = e^{-\frac{1}{2}\sum_{\mathbf{k}} \coth (\beta\omega_{\mathbf{k}}/2) |\delta\alpha_{\mathbf{k},mn}|^{2}}$. The above implies  that  in the transformed frame a localised electronic excitation fully displaces each mode of the harmonic environment, while the environment both renormalizes (see Eq. (\ref{eq:h0})) and causes fluctuations of the electronic couplings (see Eq.(\ref{eq:polaronH})). Notice that in the limit that all sites couple identically to the common bath we have that $\alpha_{\mathbf{k},m} = \alpha_{\mathbf{k}}$ and $\delta \alpha_{\mathbf{k},mn} = 0$ for all $m$ and $n$. Hence, all renormalisation factors $\beta_{mn}$ evaluate to unity, while the bath operators $\tilde{B}_{mn}$ evaluate to the null operator. Therefore, for fully correlated fluctuations we find, unsurprisingly, that the dynamics of a single-excitation is fully decoupled from the bath. 

To simplify further analysis, we now move into a basis in which $\tilde{H}_{0}$ is diagonal, denoted as the renormalised exciton basis and labeled with greek letters throughout the paper: $\tilde{H}_{0} |{\alpha}\rangle = \epsilon_{\alpha} |{\alpha}\rangle$. We may then express the original system operators in this new basis: $\sigma_{m}^{+} = \sum_{\alpha} u_{m \alpha} \sigma_{\alpha}^{+}$ where $\sigma_{\alpha}^{\dagger} = |{\alpha}\rangle\langle 0|$ and $u_{m\alpha}=\langle {\alpha}| m\rangle$. The polaron-transformed Hamiltonian within this new basis becomes
\begin{eqnarray}
\tilde{H}_{0}&{}={}& \sum_{\alpha} \epsilon_{\alpha} \sigma_{\alpha}^{+} \sigma_{\alpha}^{-} + \sum_{\mathbf{k}} \omega_{\mathbf{k}} b_{\mathbf{k}}^{\dagger} b_{\mathbf{k}},\\
\tilde{H}_{I}&{}={}& \sum_{\alpha \beta}\sum_{\langle m,n \rangle} (V_{mn} \tilde{B}_{mn} u_{m\alpha} u_{n\beta}^{*} \sigma_{\alpha}^{+} \sigma_{\beta}^{-} \nonumber\\ 
&&\:{+}V_{mn} \tilde{B}_{mn}^{\dagger} u_{m\alpha}^{*} u_{n\beta} \sigma_{\beta}^{+} \sigma_{\alpha}^{-}). 
\end{eqnarray}
Moving into the interaction picture defined with respect to $\tilde{H}_{0}$ is straightforward, and we obtain the interaction Hamiltonian
\begin{eqnarray}\label{Hint}
\tilde{H}_{I}(t) &{}={}& \sum_{\alpha \beta}\sum_{\langle m,n \rangle} \Big(V_{mn} \tilde{B}_{mn}(t) u_{m\alpha} u_{n\beta}^{*} \sigma_{\alpha}^{+} \sigma_{\beta}^{-} e^{i (\epsilon_{\alpha} - \epsilon_{\beta}) t} \nonumber\\ 
&&~~~~~~~~{+}V_{mn} \tilde{B}_{mn}^{\dagger}(t) u_{m\alpha}^{*} u_{n\beta}\sigma_{\beta}^{+} \sigma_{\alpha}^{-} e^{i (\epsilon_{\beta} - \epsilon_{\alpha}) t}\Big),\\
&{}={}&\sum_{\alpha \beta} \left(S_{\alpha \beta}(t) \otimes B_{\alpha \beta}(t) + S_{\alpha \beta}^{\dagger}(t) \otimes B_{\alpha \beta}^{\dagger}(t)\right),\nonumber\\ \label{Hint2}
\end{eqnarray}
where we have defined new system and bath operators
\begin{eqnarray}
S_{\alpha \beta}(t) &{}={}& \sigma_{\alpha}^{+} \sigma_{\beta}^{-} e^{i \epsilon_{\alpha\beta} t}, \nonumber\\
B_{\alpha \beta}(t) &{}={}& \sum_{\langle m,n\rangle} V_{mn} u_{m\alpha} u_{n\beta}^{*} \tilde{B}_{mn}(t),
\end{eqnarray}
with $\epsilon_{\alpha\beta} = \epsilon_{\alpha}-\epsilon_{\beta}$ and $\tilde{B}_{mn}(t) = e^{i \tilde{H}_{0} t} \tilde{B}_{mn} e^{-i \tilde{H}_{0} t}$. We therefore have a simple form for the polaron-frame interaction Hamiltonian $\tilde{H}_{I}(t)$, which will be treated as a perturbation in the master equation derivation that follows. 

\subsection{Time-Local Master Equation}

The time-convolutionless and the Nakajima-Zwanzig master equations are, respectively,  time-local and time-non-local perturbation expansions to describe non-Markovian dynamical evolutions \cite{breuer07}.  However, there are several reasons one can chose the time-local expansion over its non-local counterpart. As is illlustrated by Vacchini and Breuer \cite{vacchini10}, these two approaches have different ranges of validity: while the time-convolutionless expansion breaks down at finite time in the strong coupling limit, the Nakajima-Zwanzig approach does not necesarily preserve positivity if restricted to second order. This suggests that if one is able to identify the parameter space of weak system-bath coupling (as we do in our case), a second-order time-convolutionless perturbation scheme may be appropiate appropriate.  Furthermore, it is well known that time-local approaches are often simpler. This is particularly important in our case given that the polaron treatment for many-sites is already quite involved.  Most importantly, however, current research is helping to clarify that time-nonlocal and time-local approaches to non-Markovian dynamics are complementary rather than opposed. For instance, Chru\'sci\'nski and Kossakowski \cite{chruscinski10} show the retarded time integration in a time-nonlocal formalism can be mapped onto a local- time dynamics with a generator that has a strong dependence on the starting point $t_0$ . Because of the above, in this work we consider  a time-convolutionless projector operator expansion to derive a second-order, time-local master equation in the polaron representation.

\subsubsection{Projection Operator Formalism}

Having transformed our Hamiltonian into the polaron frame, we now wish to derive a time-local master equation governing the reduced dynamics of our multichromophoric excitation under the influence of the harmonic environment. In order to do so, we follow the time-local projection operator formalism (as given, for example, in Breuer and Petruccione \cite{breuer07}). In brief,  we define a projection super-operator $\mathcal{P}$ as 
\begin{equation}
\chi \rightarrow \mathcal{P}\chi = {\rm tr}_{B}\{\chi\} \otimes \rho_{\rm ref}
\end{equation}
which projects onto the relevant part of the combined system-environment density matrix $\chi$, such that $\mathcal{P}\chi$ gives the complete information required to reconstruct the reduced density matrix of the open system. Here, $\rho_{\rm ref}$ denotes a fixed (arbitrary) reference state of the environment, commonly chosen to be the thermal equilibrium state. The complementary super-operator $\mathcal{Q}$ is also defined, through $\mathcal{Q}\chi = \chi - \mathcal{P}\chi$, which projects onto the irrelevant part of the density matrix. 

Suppose the dynamics of the combined system plus bath is governed by a Hamiltonian of the general form
\begin{equation}
H = H_{0} + a H_{I},
\end{equation}
where $H_{0}$ determines the uncoupled time evolution of the system and environment, $H_{I}$ describes the system-environment interactions, and $a$ denotes a dimensionless expansion parameter. By applying the projection operators to the interaction-picture Liouville equation
\begin{equation}
\frac{d \chi(t)}{dt} = -ia [H_{I}(t), \chi(t)] = a \mathcal{L}(t) \chi(t),
\end{equation}
we may derive an exact time-convolutionless master equation for the relevant part of the density matrix of the form
\begin{equation}\label{TCL}
\frac{d}{dt} \mathcal{P} \chi(t) = \mathcal{R}(t)\chi(t) + \mathcal{I}(t) \chi(t_{0}),
\end{equation}
provided neither $a$ nor $t-t_0$ become too large, where $t_0$ is the initial time. To second order in the expansion parameter $a$, the homogeneous term $\mathcal{R}(t)$ and inhomogeneous term $\mathcal{I}(t)$ can be given as
\begin{eqnarray}
\mathcal{R}(t) &=& a^2 \int_{0}^{t} ds ~\mathcal{P} \mathcal{L}(t) \mathcal{L}(s)  \mathcal{P},\\
\mathcal{I}(t) &=& a \mathcal{P} \mathcal{L}(t) \mathcal{Q} + a^2 \int_{0}^{t} ds ~\mathcal{P} \mathcal{L}(t) \mathcal{L}(s) \mathcal{Q}, 
\end{eqnarray}
respectively, where we set $t_0=0$ and have imposed the condition that $\mathcal{P} \mathcal{L}(t)\mathcal{P}=0$.
Here, the inhomogeneous term $\mathcal{I}(t)$ is non-zero only when the initial bath state differs from the reference state $\rho_{\rm ref}$. In our case, we shall see that this corresponds to a {\it non-equilibrium} preparation of the initial environmental state within the polaron frame. 

\subsubsection{Homogeneous Term}

Let us 
now look at the homogeneous term of our master eqaution for the polaron frame interaction Hamiltonian [Eq.~(\ref{Hint})]. 
To obtain the reduced dynamics for the many-site system, 
$\tilde{\rho}(t)$, we trace over the bath degrees of freedom to arrive at:
\begin{eqnarray}
\mathcal{R}(t)\tilde{\rho}(t) &{}={}& \textrm{tr}_{B} \Big \{ a^2 \int_{0}^{t} ds~\mathcal{P}\mathcal{L}(t)\mathcal{L}(s)\mathcal{P} \tilde{\chi} \Big\} \nonumber\\
&{}={}& - a^2 \int_{0}^{t} ds~\textrm{tr}_{B} \Big \{ \big[ \mathcal{L}(t) \mathcal{L}(s) \tilde{\rho}(t){\rho}_{\rm ref} \big] \Big \} \nonumber\\
&{}={}& - \int_{0}^{t} ds~\textrm{tr}_{B} \Big \{ \big[ \tilde{H}_{I}(t), [ \tilde{H}_{I}(s), \tilde{\rho}(t){\rho}_{\rm ref} ] \big] \Big\}. \nonumber \\
\end{eqnarray}
We shall now take our bath reference state ${\rho}_{\rm ref}$ to be the thermal equilibrium state within the polaron frame i.e.  ${\rho}_{\rm ref}\equiv  \tilde{\rho}_{B}$. On substituting in the interaction Hamiltonian given in Eq.~(\ref{Hint2}), we obtain

\begin{eqnarray}
R(t)\tilde{\rho}(t)&{} ={}& - \sum_{\alpha \beta \mu \nu} \Big( \Gamma_{\alpha\beta,\mu\nu}^{(1)}(t) e^{i \epsilon_{\alpha\beta} t}  \big[S_{\alpha\beta}, S_{\mu\nu} \tilde{\rho}(t) \big] \nonumber\\ 
&& ~~~~~~~~~{+} \Gamma_{\alpha\beta,\mu\nu}^{(2)}(t) e^{i \epsilon_{\beta\alpha} t} \big[S_{\alpha\beta}^{\dagger}, S_{\mu\nu} \tilde{\rho}(t) \big] \nonumber\\ 
&& ~~~~~~~~~{+} \Gamma_{\alpha\beta,\mu\nu}^{(3)}(t) e^{i \epsilon_{\alpha\beta} t} \big[ S_{\alpha\beta}, S_{\mu\nu}^{\dagger} \tilde{\rho}(t) \big]  \nonumber\\
&& ~~~~~~~~~{+} \Gamma_{\alpha\beta,\mu\nu}^{(4)}(t) e^{i \epsilon_{\beta\alpha} t} \big[ S_{\alpha\beta}^{\dagger}, S_{\mu\nu}^{\dagger} \tilde{\rho}(t) \big] \nonumber\\ 
&& ~~~~~~~~~{+} {\rm h.c.} \Big).
\label{eq:R}
\end{eqnarray}
Here, we have defined the time-dependent rates
\begin{eqnarray}
\Gamma_{\alpha\beta,\mu\nu}^{(1)}(t) &{}={}& \int_{0}^{t} ds ~e^{i\epsilon_{\mu\nu}s} ~C_{\alpha\beta,\mu\nu}^{(1)}(t-s), \nonumber\\
\Gamma_{\alpha\beta,\mu\nu}^{(2)}(t) &{}={}& \int_{0}^{t} ds ~e^{i\epsilon_{\mu\nu}s} ~C_{\alpha\beta,\mu\nu}^{(2)}(t-s), \nonumber\\
\Gamma_{\alpha\beta,\mu\nu}^{(3)}(t) &{}={}& \int_{0}^{t} ds ~e^{i\epsilon_{\nu\mu}s} ~C_{\alpha\beta,\mu\nu}^{(3)}(t-s), \nonumber\\
\Gamma_{\alpha\beta,\mu\nu}^{(4)}(t) &{}={}& \int_{0}^{t} ds ~e^{i\epsilon_{\nu\mu}s} ~C_{\alpha\beta,\mu\nu}^{(4)}(t-s),
\label{eq:homogeneousrates}
\end{eqnarray}
with corresponding two-time bath correlation functions 
\begin{eqnarray}
C_{\alpha\beta,\mu\nu}^{(1)}(t-s) &=& \sum_{\langle mn \rangle} \sum_{\langle pq \rangle} \mathcal{U}_{\alpha\beta\mu\nu}^{mnpq} ~\langle \tilde{B}_{mn}(t) \tilde{B}_{pq}(s) \rangle, \nonumber\\ 
C_{\alpha\beta,\mu\nu}^{(2)}(t-s) &=& \sum_{\langle mn \rangle} \sum_{\langle pq \rangle}\mathcal{U}_{\alpha\beta\mu\nu}^{mnpq} ~\langle \tilde{B}_{mn}^{\dagger}(t) \tilde{B}_{pq}(s) \rangle, \nonumber\\ 
C_{\alpha\beta,\mu\nu}^{(3)}(t-s) &=& \sum_{\langle mn \rangle} \sum_{\langle pq \rangle}\mathcal{U}_{\alpha\beta\mu\nu}^{mnpq} ~\langle \tilde{B}_{mn}(t) \tilde{B}_{pq}^{\dagger}(s) \rangle,\nonumber\\ 
C_{\alpha\beta,\mu\nu}^{(4)}(t-s) &=& \sum_{\langle mn \rangle} \sum_{\langle pq \rangle}\mathcal{U}_{\alpha\beta\mu\nu}^{mnpq} ~\langle \tilde{B}_{mn}^{\dagger}(t) \tilde{B}_{pq}^{\dagger}(s) \rangle,
\end{eqnarray}
where $\mathcal{U}_{\alpha\beta\mu\nu}^{mnpq} = V_{mn} V_{pq} u_{m\alpha} u_{n\beta} u_{p\mu} u_{q \nu}$, and $\langle ... \rangle$ denotes the average with respect to the reference state of the bath $\tilde{\rho}_{B}$. 
The derivation of the explicit form of the two-time correlation functions is rather involved, so we show the exact expressions in Appendix \ref{sec:a1}. 

\subsubsection{Inhomogeneous term}

If the initial bath state within the polaron frame differs from the reference bath state  $\tilde{\rho}_{B}$ i.e. it is a non-thermal equilibrium state within the polaron frame, then we must also account for the inhomogeneous term in the master equation (see Eq.~(\ref{TCL})). 

Let us consider a general separable initial state in the {\it lab} frame (i.e. prior to polaron tranformation): $\chi(0) = \sum_{ij} \rho_{ij}(0) \sigma_{i}^{+} \sigma_{j}^{-} \otimes \rho_{B}$, where $\rho_{B}$ denotes the thermal equilibrium bath state in the lab frame. Transforming into the polaron frame we find the initial state
\begin{equation}
\tilde{\chi}(0) = \sum_{ij} \tilde{\rho}_{ij}(0) \sigma_{i}^{+} \sigma_{j}^{-} \prod_{\mathbf{k}} \beta_{ij}^{-1} D(\alpha_{\mathbf{k},i}) \tilde{\rho}_{B} D(-\alpha_{\mathbf{k},j}).
\end{equation}
Here $\tilde{\rho}_{ij}(0) = \beta_{ij} \rho_{ij}(0)$  and denotes the $ij$'th element of the initial system density operator in the polaron frame and $D(\alpha_{\mathbf{k},j})=e^{\alpha_{\mathbf{k},j} b_{\mathbf{k}}^{\dagger} - \alpha_{\mathbf{k},j }^{*} b_{\mathbf{k}}}$ is the bath displacement operator of mode $\mathbf{k}$ due to interaction with site $j$.

The irrelevant part of the total system-bath density matrix at time zero is then given by
\begin{eqnarray}
\mathcal{Q}\tilde{\chi}(0) &{}={}& \sum_{ij} \tilde{\rho}_{ij}(0) \sigma_{i}^{+} \sigma_{j}^{-}\prod_{\mathbf{k}} \Big( \beta_{ij}^{-1} D(\alpha_{\mathbf{k},i}) \tilde{\rho}_{B} D(-\alpha_{\mathbf{k},j}) - \tilde{\rho}_{B} \Big) \nonumber\\
&=& \sum_{ij} \tilde{\rho}_{ij}(0) \sigma_{i}^{+} \sigma_{j}^{-}Q_{ij} \tilde{\rho}_{B} .
\end{eqnarray}
Notice that we have defined $Q_{ij}\tilde{\rho}_{B} $ as the state accounting for the difference between the displaced bath and the bath thermal equilibrium in the polaron frame. We are now in a position to evaluate the inhomogeneous term; tracing over the bath we obtain 
\begin{eqnarray}
\mathcal{I}(t) \tilde{\rho}(0) &=& a ~ \mathrm{tr}_{B} \big \{ \mathcal{P} \mathcal{L}(t) \mathcal{Q} \tilde{\chi}(0) \big\} \nonumber\\ && + a^{2} \int_{0}^{t} ds ~ \mathrm{tr}_{B} \big\{ \mathcal{P} \mathcal{L}(t) \mathcal{L}(s) \mathcal{Q} \tilde{\chi}(0) \big\}.
\label{eq:Itotal}
\end{eqnarray}
Let us consider each term separately. 
On substituting in the form of the interaction Hamiltonian to the term first order in $a$, we obtain 
\begin{eqnarray}
\mathcal{I}_{1}(t) \tilde{\rho}(0) &=& -i \mathrm{tr}_{B} \big\{ [\tilde{H}_{I}(t), \mathcal{Q} \tilde{\chi}(0) ] \big\} \nonumber\\
&=& -i \sum_{\alpha \beta} \sum_{ij} \big (\tilde{\rho}_{ij}(0) \Upsilon_{ij,\alpha\beta}(t) e^{i\epsilon_{\alpha\beta}t} [S_{\alpha\beta}, \sigma_{i}^{+} \sigma_{j}^{-} ] \nonumber\\ && ~~~~~~~~~~~~~~~~~ + h.c. \Big)
\label{eq:I1}
\end{eqnarray}
Here we have introduced the rate  $\Upsilon_{ij,\alpha\beta} (t) = \sum_{\langle m,n\rangle} V_{mn} u_{m\alpha} u_{n\beta}^{*} \langle \tilde{B}_{mn}(t) \rangle_{\mathcal{Q}_{ij}\tilde{\rho}_{B}}$ (see Appendix~\ref{sec:a2}), where $\langle ... \rangle_{\mathcal{Q}_{ij}\tilde{\rho}_{B}}$ denotes a thermal average with respect to $\mathcal{Q}_{ij}\tilde{\rho}_{B}$. 

The second order in $a$ term of the inhomogeneous super-operator is

\begin{eqnarray}
\mathcal{I}_{2}(t) \tilde{\rho}(0) &{}={}& a^{2} \int_{0}^{t} ds ~ \mathrm{tr}_{B} \big \{ \mathcal{P} \mathcal{L}(t) \mathcal{L}(s) \mathcal{Q} \tilde{\chi}(0) \big\} \nonumber\\
&{}={}& - \int_{0}^{t} ds ~ \mathrm{tr}_{B} \big \{ [\tilde{H}_{I}(t), [\tilde{H}_{I}(s) , \mathcal{Q} \tilde{\chi}(0) ] ] \big \},\nonumber\\ 
\end{eqnarray}
Substituting in the interaction Hamiltonian, we obtain:
\begin{eqnarray}
\mathcal{I}_{2}(t) \tilde{\rho}(0) &=& - \sum_{\substack{ ij \\ \alpha\beta\mu\nu}} \Big( \tilde{\rho}_{ij}(0) \Xi_{ij,\alpha\beta,\mu\nu}^{(1)} (t) e^{i \epsilon_{\alpha\beta} t} [S_{\alpha\beta} , S_{\mu\nu} \sigma_{i}^{+} \sigma_{j}^{-} ] \nonumber\\
&& ~~~~~~~~ + \tilde{\rho}_{ij}(0) \Xi_{ij,\alpha\beta,\mu\nu}^{(2)} (t) e^{i \epsilon_{\beta\alpha} t} [S_{\alpha\beta}^{\dagger} , S_{\mu\nu} \sigma_{i}^{+} \sigma_{j}^{-} ] \nonumber\\
&& ~~~~~~~~ + \tilde{\rho}_{ij}(0) \Xi_{ij,\alpha\beta,\mu\nu}^{(3)} (t) e^{i \epsilon_{\alpha\beta} t} [S_{\alpha\beta} , S_{\mu\nu}^{\dagger} \sigma_{i}^{+} \sigma_{j}^{-} ] \nonumber\\
&& ~~~~~~~~ + \tilde{\rho}_{ij}(0) \Xi_{ij,\alpha\beta,\mu\nu}^{(4)} (t) e^{i \epsilon_{\beta\alpha} t} [S_{\alpha\beta}^{\dagger} , S_{\mu\nu}^{\dagger} \sigma_{i}^{+} \sigma_{j}^{-} ] \nonumber\\
&& ~~~~~~~~ + h.c. \Big),
\label{eq:I2}
\end{eqnarray}
where we have defined the rates:
\begin{eqnarray}
\Xi_{ij,\alpha\beta,\mu\nu}^{(1)} (t) &=& \int_{0}^{t} ds~ e^{i\epsilon_{\mu\nu} t} D_{ij,\alpha\beta,\mu\nu}^{(1)}(t-s), \nonumber\\
\Xi_{ij,\alpha\beta,\mu\nu}^{(2)} (t) &=& \int_{0}^{t} ds~ e^{i\epsilon_{\mu\nu} t} D_{ij,\alpha\beta,\mu\nu}^{(2)}(t-s), \nonumber\\
\Xi_{ij,\alpha\beta,\mu\nu}^{(3)} (t) &=& \int_{0}^{t} ds~ e^{i\epsilon_{\nu\mu} t} D_{ij,\alpha\beta,\mu\nu}^{(3)}(t-s), \nonumber\\
\Xi_{ij,\alpha\beta,\mu\nu}^{(4)} (t) &=& \int_{0}^{t} ds~ e^{i\epsilon_{\nu\mu} t} D_{ij,\alpha\beta,\mu\nu}^{(4)}(t-s),
\end{eqnarray}
with bath correlation functions 
\begin{eqnarray}
D_{ij,\alpha\beta,\mu\nu}^{(1)}(t-s) = \sum_{\langle mn \rangle} \sum_{\langle pq \rangle} \mathcal{U}_{\alpha\beta\mu\nu}^{mnpq} \langle \tilde{B}_{mn}(t) \tilde{B}_{pq}(s)\rangle_{Q_{ij}\tilde{\rho}_{B}}, \nonumber\\ 
D_{ij,\alpha\beta,\mu\nu}^{(2)}(t-s) = \sum_{\langle mn \rangle} \sum_{\langle pq \rangle} \mathcal{U}_{\alpha\beta\mu\nu}^{mnpq} \langle \tilde{B}_{mn}^{\dagger}(t) \tilde{B}_{pq}(s) \rangle_{Q_{ij}\tilde{\rho}_{B}}, \nonumber\\ 
D_{ij,\alpha\beta,\mu\nu}^{(3)}(t-s) = \sum_{\langle mn \rangle} \sum_{\langle pq \rangle} \mathcal{U}_{\alpha\beta\mu\nu}^{mnpq} \langle \tilde{B}_{mn}(t) \tilde{B}_{pq}^{\dagger}(s)  \rangle_{Q_{ij}\tilde{\rho}_{B}}, \nonumber\\ 
D_{ij,\alpha\beta,\mu\nu}^{(4)}(t-s) = \sum_{\langle mn \rangle} \sum_{\langle pq \rangle} \mathcal{U}_{\alpha\beta\mu\nu}^{mnpq} \langle \tilde{B}_{mn}^{\dagger}(t) \tilde{B}_{pq}^{\dagger}(s) \rangle_{Q_{ij}\tilde{\rho}_{B}}. \nonumber\\
\end{eqnarray}

\noindent Once again, explicit forms for these correlation functions are presented in Appendix \ref{sec:a2}.

\subsection{Lab Frame Dynamics}

The master equation derived above gives the dynamics of the reduced density matrix for the electronic system within the polaron frame. However, we are interested in the excitation dynamics in the original untransformed lab frame. To calculate the correct transformation from polaron to lab frame, consider the Schr\"odinger picture system-bath density operator in the polaron frame  $\tilde{\chi}(t) = e^{S} \chi(t) e^{-S}$. Inverting this expression and using the identity $\mathcal{P}+\mathcal{Q} = I$, we may write the lab frame combined density operator as:

\begin{equation}
\chi(t) = e^{-S} \mathcal{P} \tilde{\chi}(t) e^{S} + e^{-S} \mathcal{Q} \tilde{\chi}(t) e^{S}.
\end{equation}

\noindent The expectation value of a system observable $A$ in the lab frame is given by

\begin{eqnarray}
\langle A \rangle &=& \textrm{tr}_{S+B} \{A \chi(t)\} \nonumber\\
&=& \textrm{tr}_{S+B} \{e^{S} A  e^{-S} \mathcal{P} \tilde{\chi}(t)\}  + \textrm{tr}_{S+B} \{e^{S} A  e^{-S} \mathcal{Q} \tilde{\chi}(t) \} \nonumber\\
&=& \langle A \rangle_{\rm{rel}} +  \langle A \rangle_{\rm{irrel}}\ .
\end{eqnarray}
From the definition of the projection operator, the first term $ \langle A \rangle_{rel}$ is trivial to evaluate:

\begin{eqnarray}
\langle A \rangle_{\rm{rel}} &=&\textrm{tr}_{S+B} \{e^{S} A  e^{-S} \mathcal{P} \tilde{\chi}(t) \} \nonumber\\
&=& \textrm{tr}_{S+B} \{e^{S} A e^{-S} \tilde{\rho}(t) \otimes \tilde{\rho}_{B} \} \nonumber\\
&=& \textrm{tr}_{S} \{ \tilde{A} \tilde{\rho}(t) \},
\end{eqnarray}
where we have defined the transformed observable $\tilde{A} = \textrm{tr}_{B}\{ e^{S} A e^{-S} \tilde{\rho}_{B}\}$. Since this contribution depends entirely on the relevant dynamics, we have defined it as the {\it relevant contribution} to the expected value of $A$. 

To evaluate the second term, $\langle A \rangle_{\rm{irrel}}$, we require knowledge of the dynamics of the irrelevant part of the density matrix. Breuer and Petruccione \cite{breuer07} show that the irrelevant part at an arbitrary time $t$ can in principle be determined from the knowledge of both the relevant part $\mathcal{P}\tilde{\chi}(t)$ and the initial condition $\mathcal{Q}\tilde{\chi}(0)$. Importantly, in our case, both these quantities are known. Therefore, we may formally write the irrelevant part as $\mathcal{Q}\tilde{\chi}(t) = \mathcal{S}(t) \mathcal{P}\tilde{\chi}(t) + \mathcal{T}(t) \mathcal{Q}\tilde{\chi}(0)$. The resulting {\it irrelevant contribution}  to the system operator expectation value is 

\begin{eqnarray}
\langle A \rangle_{\rm{irrel}} &=& \textrm{tr}_{S+B} \{e^{S} A  e^{-S} \mathcal{Q} \tilde{\chi}(t) \}\\
&=& \textrm{tr}_{S+B} \{e^{S} A  e^{-S} \mathcal{S}(t) \mathcal{P} \tilde{\chi}(t) \} \nonumber\\ 
&& + \textrm{tr}_{S+B} \{ e^{S} A  e^{-S} \mathcal{T}(t) \mathcal{Q} \tilde{\chi}(0) \}. \nonumber\\
\end{eqnarray}

\noindent Up to second order in the coupling $a$, the super-operators $\mathcal{S}(t)$ and $\mathcal{T}(t)$ are given by:

\begin{eqnarray}
\mathcal{S}(t) &=& a \int_{0}^{t} ds \mathcal{L}(s) + a^{2} \int_{0}^{t} ds \int_{0}^{s} ds' \mathcal{Q} \mathcal{L}(s) \mathcal{L}(s'), \nonumber\\
\mathcal{T}(t) &=& 1 + a \int_{0}^{t} ds \mathcal{Q} \mathcal{L}(s) + a^{2} \int_{0}^{t} ds \int_{0}^{s} ds' \mathcal{Q} \mathcal{L}(s) \mathcal{Q} \mathcal{L}(s') \nonumber\\
&& ~~~~~~~~~~~ - a^{2} \int_{0}^{t} ds \int_{0}^{s} ds' \mathcal{Q} \mathcal{L}(s') \mathcal{P} \mathcal{L}(s).
\end{eqnarray}

In general, the exact forms and calculations of the irrelevant contributions to expectation values are extremely involved and for simplicity here we restict ourselves to operators for which such terms evaluate to zero as it is explained below. However, it is worth noting that to zeroth order in the coupling parameter $a$ the irrelevant contribution becomes $\langle A \rangle_{\rm{irrel}}= \textrm{tr}_{S+B} \{e^{S} A  e^{-S} \mathcal{Q} \tilde{\chi}(0) \} $, which can be used to evaluate approximate expectation values of system operators that do not commute with the polaron transformation $S$. Specifically, the expectation value of an observable in the lab frame including just the zeroth order term for the irrelevant contribution reads

\begin{eqnarray}
\langle A \rangle =\textrm{tr}_{S} \{\tilde{A} \tilde{\rho}(t) \} + \textrm{tr}_{S} \{A \rho(0) \} - \textrm{tr}_{S} \{\tilde{A} \tilde{\rho}(0) \}.
\label{eq:A}
\end{eqnarray}

In the case of system operators commuting with the polaron transformation $S$, such as the $m-$th site population operator  $\sigma_{m}^{+} \sigma_{m}^{-}$,  the irrelevant contribution vanishes and the expected value in the lab frame is entirely determined by the relevant contribution.  In other words, site populations remain unaffected during the transformation back to the lab frame. Let us demonstrate this explicitly:

\begin{eqnarray}
\langle \sigma_{m}^{+} \sigma_{m}^{-} \rangle &=& \textrm{tr}_{S+B} \{\sigma_{m}^{+} \sigma_{m}^{-} \mathcal{P} \tilde{\chi}(t) \} + \textrm{tr}_{S+B} 
\{ \sigma_{m}^{+} \sigma_{m}^{-} \mathcal{Q} \tilde{\chi}(t) \} \nonumber\\
&=& \textrm{tr}_{S} \{ \sigma_{m}^{+} \sigma_{m}^{-} \tilde{\rho}(t) \} + \textrm{tr}_{S} \big \{ \sigma_{m}^{+} \sigma_{m}^{-} ~\textrm{tr}_{B}\{\mathcal{Q} \tilde{\chi}(t) \}\big \} \nonumber\\
&=&\textrm{tr}_{S} \{\sigma_{m}^{\dagger}\sigma_{m} \tilde{\rho}(t)\}=\langle \sigma_{m}^{+} \sigma_{m}^{-} \rangle_{\rm{rel}},
\end{eqnarray}
where by definition $\textrm{tr}_{B}\{\chi\} = \textrm{tr}_{B}\{\mathcal{P}\chi\}$ and $\mathcal{P}\mathcal{Q}=0$, and therefore $\langle \sigma_{m}^{+} \sigma_{m}^{-} \rangle_{\rm{irrel}}=0$. In Sec. \ref{sec:3} we show that the key aspect of the theory can be illustrated by considering site populations.

One may attempt to compute off-diagonal operators in the site basis, i.e. $\sigma_{m}^{+} \sigma_{n}^{-}$ with $m\neq n$, with the approximation proposed in Eq.(\ref{eq:A}) which we present here for completeness.  
\begin{eqnarray}
\langle \sigma_{m}^{+} \sigma_{n}^{-} \rangle =\beta_{mn} \textrm{tr}_{S} \{\sigma_{m}^{+} \sigma_{n}^{-} \tilde{\rho}(t) \} + \textrm{tr}_{S} \{\sigma_{m}^{+} \sigma_{n}^{-} \rho(0) \} \nonumber\\ - \beta_{mn} \textrm{tr}_{S} \{\sigma_{m}^{+} \sigma_{n}^{-} \tilde{\rho}(0) \}. ~~~~~~~~~~~~~~~
\end{eqnarray}
 
Let us finish this section with a discussion of the validity of this polaron treatment.   As our approach is pertubative we expect the master equation to be valid only within certain regimes. Firstly, notice that in the absence of electronic couplings the interaction Hamiltonian $\tilde{H}_{I}$ (Eq.(\ref{eq:polaronH})) is zero and the polaron transformation exactly diagonalises the combined system-bath Hamiltonian. Therefore, we expect this perturbative treatment to be a good approximation in the limit where the magnitude of electronic couplings are small in comparison to the detunings between onsite energies, irrespective of the strength of the coupling to the bath \cite{silbey75, silbey89}. Moreover, this perturbative treatment is valid if the energy scale associated to the fluctuations of the electronic couplings represented by $\tilde{H}_{I}$ is the smallest energy scale in the system. Such fluctuations are given by \cite{jang08}
\begin{eqnarray}
\gamma_{mn} &=& V_{mn} \langle |\tilde{B}_{mn}|^{2} \rangle^{1/2} \nonumber\\ &=& V_{mn} (1 -\beta_{mn}^{2})^{1/2}.
\end{eqnarray} 

Furthermore, as mentioned before, the small polaron transformation $S$ assumes that bath modes are fully displaced by the interaction with a localised electronic excitation. For super-Ohmic spectral densities such full displacement is only valid for bath frequencies larger than the typical energy scale of the renormalized excitonic Hamiltonian \cite{silbey75, silbey89}. Notice also that for an Ohmic spectral density, i.e. $J(\omega)\propto \omega$, and independent baths for each site, the renormalization factors exhibit a well-known infra-red divergence \cite{silbey89} that leads to electronic couplings being renormalized to zero independently of the strength of the system-environment interaction. Both of these short-comings can in principle be alleviated by considering a variational-polaron approach \cite {silbey89} or by developing perturbative treatments based on alternative transformations of the system-bath interaction \cite{irene09, prior10}.

\section{Non-Markovian Dynamics}
\label{sec:3}

To illustrate the scope of this theory, we apply it to study the dynamics of a subsystem of the Fenna-Matthews-Olsen (FMO) complex. In particular, we consider the subsystem involving sites 1, 2, 3 and 4 with an electronic Hamiltonian taken from Cho \textit{et al.} \cite{cho05}. In units of $\textrm{cm}^{-1}$ this reads:

\begin{equation}
H_{FMO} = \left( 
\begin{array}{cccc} 
280 & -106 & 8 & -5 \\ 
-106 & 420 & 28 & 6  \\
8 & 28 & 0 & -62 \\
-5 & 6 & -62 & 175
\end{array} \right). 
\end{equation}

Each site is coupled to an independent bath with spectral density given by $J(\omega) = s_{0} J_{0}(\omega) + s_{H} J_{H}(\omega)$ \cite{renger02}, which has a continuous contribution  $J_{0}(\omega)$ and a contribution from a localised vibrational mode $J_{H}(\omega)$. The continuous part of the spectral density is defined as:

\begin{equation}
J_{0}(\omega) = \frac{1}{s_{1}+s_{2}} \sum_{i=1,2} \frac{s_{i}\omega^{5}}{7! 2\omega_{i}^{4}} e^{-(\omega/\omega_{i})^{1/2}}.
\label{eq:sd}
\end{equation}

\noindent The localised vibrational mode is commonly described by a delta function. This provides a simple picture for describing the coupling to localised modes and allows for analytical expressions for correlation functions, but does not necessarily represent a realistic situation. One would expect that in practice the single frequency mode is broadened by interactions with the surrounding bulk modes \cite{hausinger08,garg85}. Therefore, in this work we shall assume a broadened vibrational mode with a Lorentzian line shape: 

\begin{equation}
J_{H}(\omega) = \frac{2\omega_{H}}{\pi} \frac{\omega^{3} \epsilon}{(\omega^{2}-\omega_{H}^{2})^{2} + \epsilon^{2}\omega^{2}}.
\end{equation}

\noindent Here, the parameters for the continuous part of the spectral density are $s_{0}=0.5$, $s_{1}=0.8$, $s_{2}=0.5$, $\omega_{1} =0.0069~\textrm{meV}$ and $\omega_{2} =0.024~\textrm{meV}$. Meanwhile, the parameters for the localised mode are $s_{H} = 0.22$, $\omega_{H} = 180~\textrm{cm}^{-1}$ and broadening $\epsilon=50~\textrm{cm}^{-1}$, a value that has been chosen to be larger than the average electronic coupling strength. For all the calculations, room temperature is assumed, i.e. $\textrm{k}_{B}T=200~\textrm{cm}^{-1}$

For the FMO spectral density introduced above, the bath renormalisation factors may be written as

\begin{eqnarray}
\beta_{mn} = \textrm{exp}\Big(-\int_{0}^{\infty} d\omega \frac{J_{0}(\omega)}{\omega^{2}} \coth(\beta\omega/2)\Big) ~~~~~~~~ \nonumber\\
\times~\textrm{exp}\Big(-\int_{0}^{\infty} d\omega \frac{J_{H}(\omega)}{\omega^{2}} \coth(\beta\omega/2)\Big)
\end{eqnarray}

\noindent and the renormalised electronic Hamiltonian then evaluates to

\begin{equation}
\tilde{H}_{FMO} = \left( 
\begin{array}{cccc} 
280 & -0.107 & 0.008 & -0.005 \\ 
-0.107 & 420 & 0.028 & 0.006  \\
0.008 & 0.028 & 0 & -0.062 \\
-0.005 & 0.006 & -0.062 & 175
\end{array} \right). 
\end{equation}

\noindent The theory predics that the harmonic bath renormalises the electronic couplings to such a degree that the excitonic eigenstates are effectively localised on sites. Hence the transition frequency between the two highest energy eigenstates is set by the difference between the energies of sites 1 and 2. It is worth noting here that these strong bath-induced renormalisations predominantly arise from the continuous part of the spectral density $J_{0}(\omega)$.  For the parameters given above onsite energy gaps are all larger than electronic couplings between sites i.e. $|\epsilon_m ~ -~ \epsilon_n|> |V_{mn}|$ though electronic coupling fluctuations $\gamma_{mn}$ are comparable with $V_{mn}$. Nevertheless, all $\gamma_{mn}$ are still smaller than the characteristic frequency of $J_{0}(\omega) $ (around 200 cm$^{-1}$), which makes the present polaron treatment appropriate. 

\begin{figure}[t!]
\subfigure[]{
\includegraphics[width=90mm]{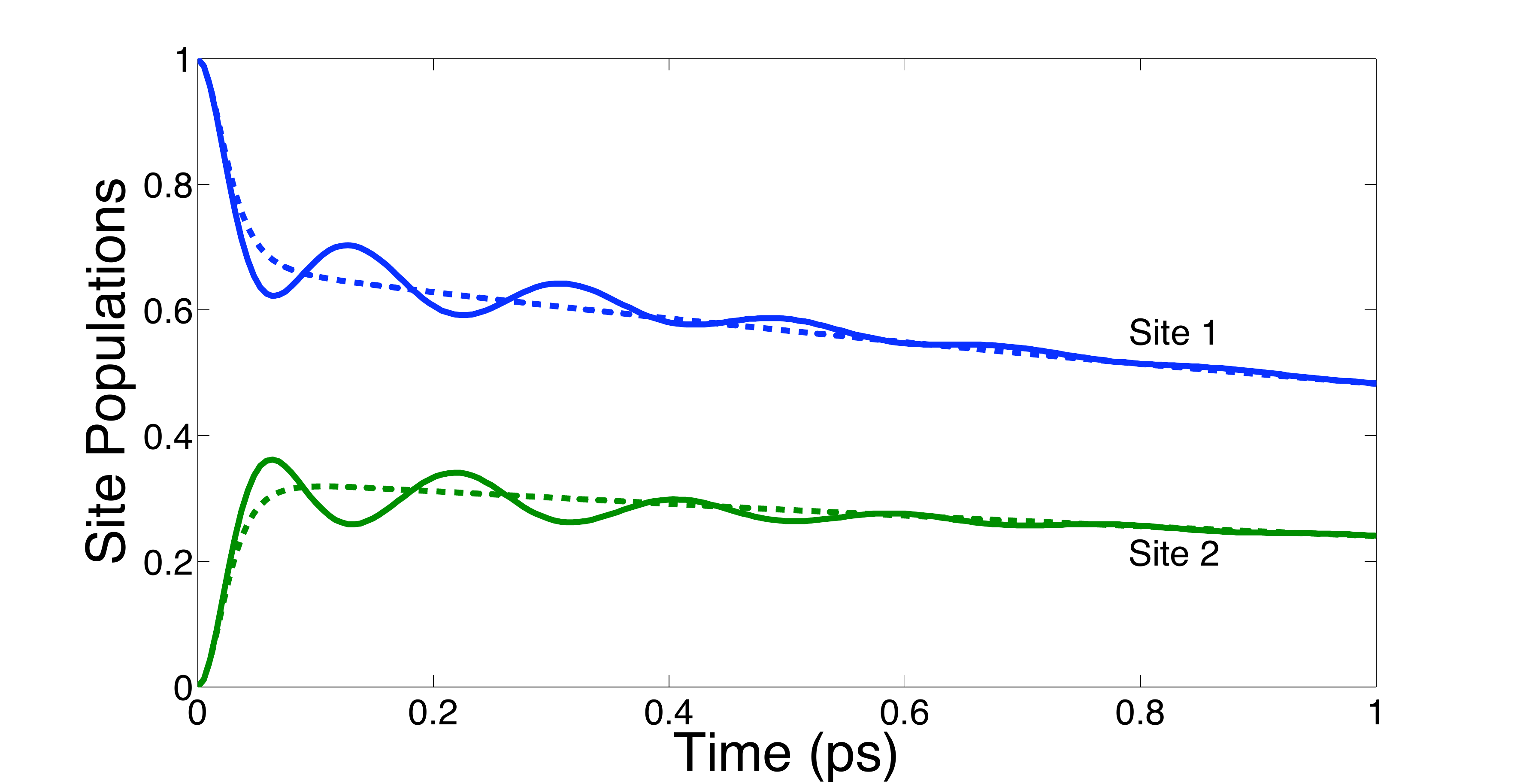}
}
\subfigure[]{
\includegraphics[width=90mm]{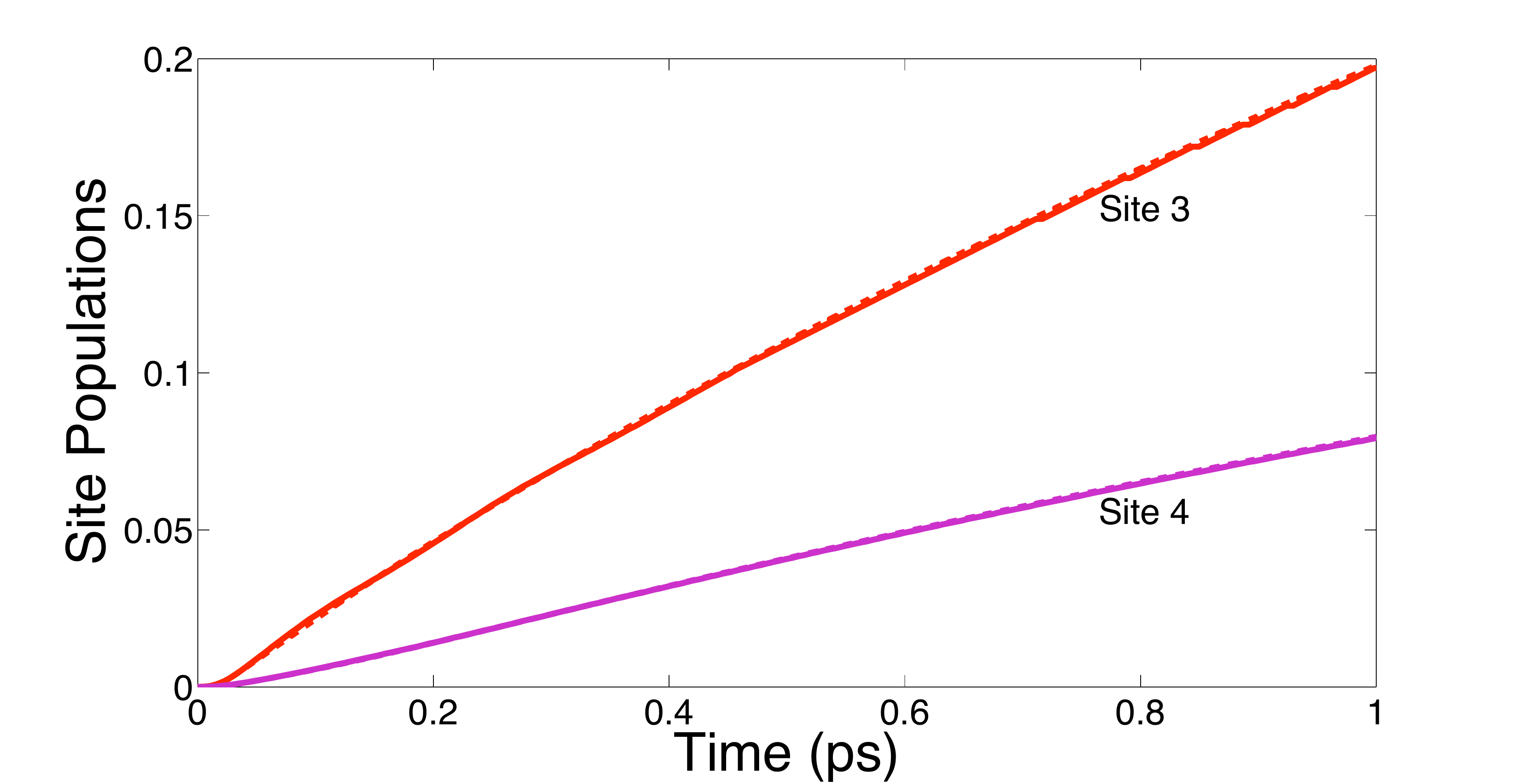}
}
\caption{Population dynamics of the FMO subsystem assuming an initial state localised on site 1. Presented are the dynamics with just the homogeneous superoperator (dashed), and the homogeneous plus inhomogeneous superoperators (solid).}
\label{fig:NMlocalised}
\end{figure}

\subsection{Non-Equilibrium Effects}

We now investigate the dynamics of the FMO subsystem as predicted by the non-Markovian polaron theory. We begin by considering an initial state localised on site 1. Figure \ref{fig:NMlocalised} shows the dynamics of populations of the four sites and compare the effects of the different terms of the non-Markovian polaron master equation. In the presence of the homogeneous superoperator term alone, all four populations evolve monotonically with no discernable oscillatory dynamics. On including the inhomogeneous superoperator in the polaron master equation, we see remarkably the emergence of oscillatory dynamics in the populations of sites 1 and 2. This behaviour is long lived, lasting upto 600~fs. Beyond the 600~fs timescale we see no variation between the dynamics considering solely the homogeneous term and the full polaron master equation. Therefore, one can conclude that the inhomogeneous terms, which describe non-equilibrium bath effects, have a profound effect at short times allowing for the emergence of oscillatory dynamics in agreement with previous works \cite{ishizaki09c}. 

\begin{figure}[t!]
\subfigure[]{
\includegraphics[width=90mm]{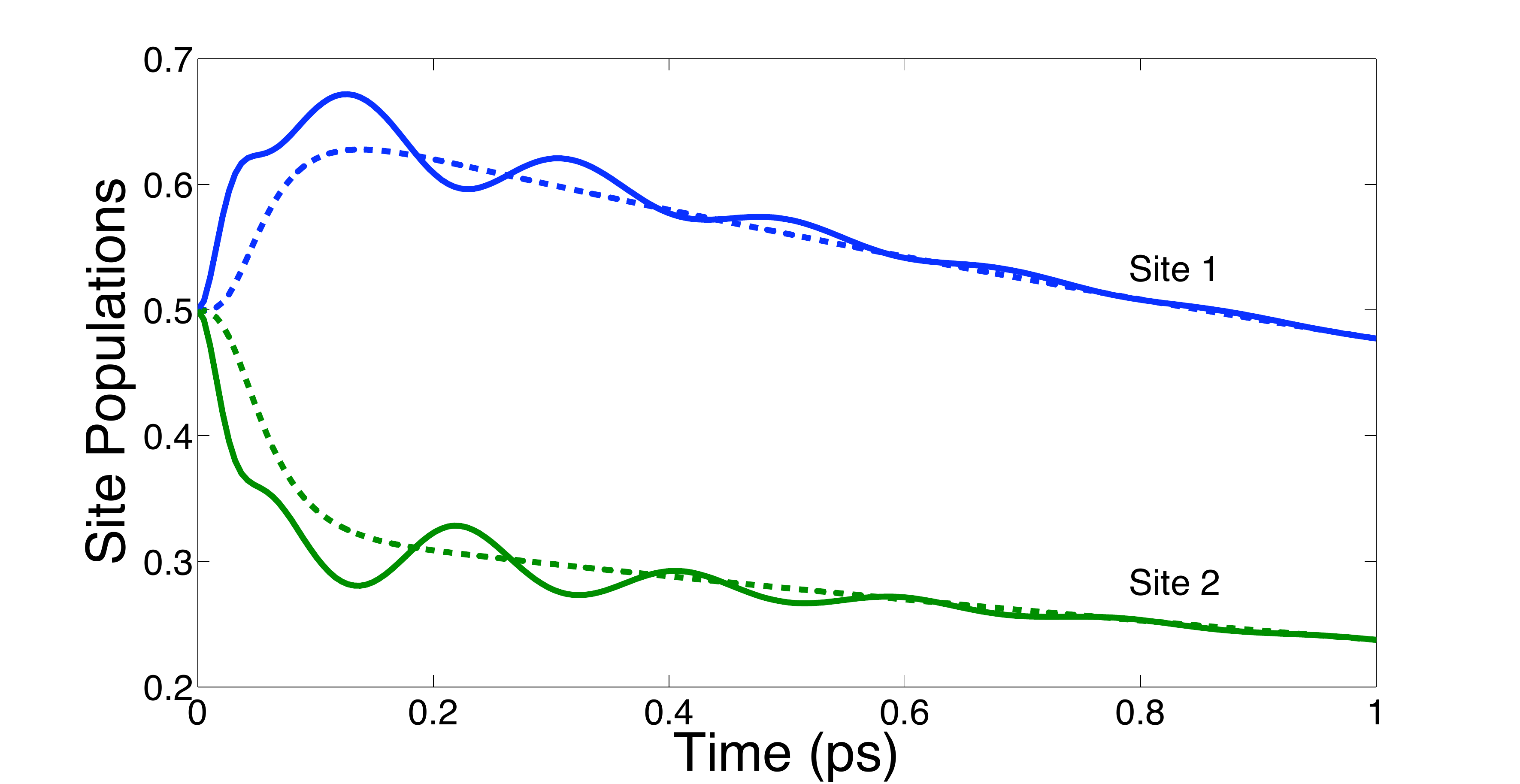}
}
\subfigure[]{
\includegraphics[width=90mm]{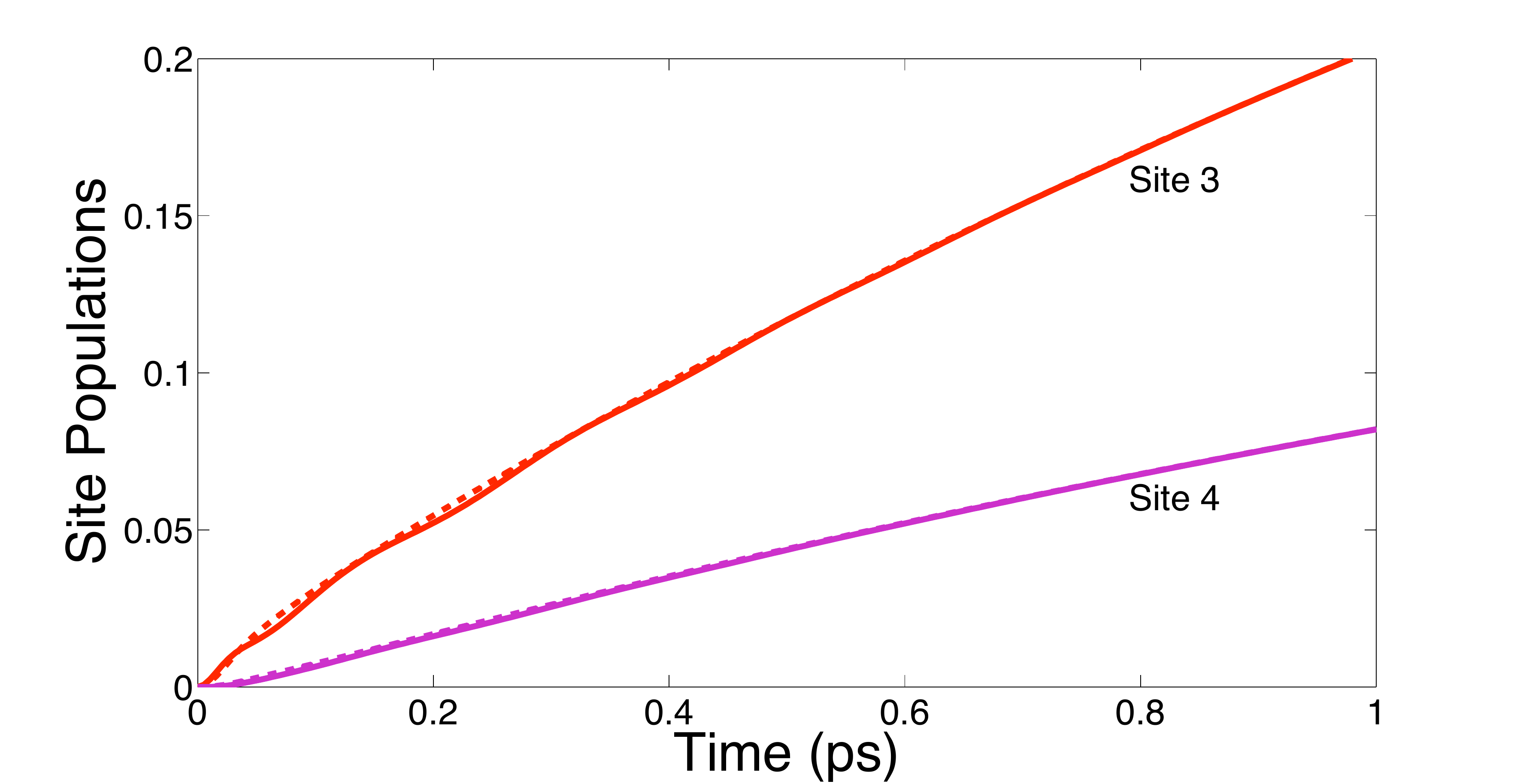}
}
\caption{Population dynamics of the FMO subsystem assuming an initial state consisting of a symmetric superposition of sites 1 and 2. Presented are the dynamics with just the homogeneous superoperator (dashed), and the homogeneous plus full inhomogeneous superoperators (solid).}
\label{fig:NMdelocalised}
\end{figure}

\subsection{Delocalised Initial states}

We now consider an electronic excitation  initially  delocalised over a number of sites while being separable with the thermal equilibrium bath in the lab frame. Upon transformation into the polaron frame,  this state maps onto an initially  correlated system-bath state.  Assuming an electronic excitation symmetrically delocalised between sites 1 and 2, figure  \ref{fig:NMdelocalised}, depicts the population dynamics of each site, comparing the evolutions given by the homogeneous term versus the full the polaron master equation. In the presence of the homogeneous term alone, the populations evolve incoherently as in the case of localised excitation. Upon the inclusion of the inhomogeneous terms we once again see the emergence of well-defined, long lasting oscillations. Interestingly, along with the coherent oscillations in the dynamics of sites 1 and 2, we are also able to observe subtle oscillatory behaviour in the population of site 3. These oscillations can be seen to decay over the same 600~fs timescale observed in the dynamics for a localised initial state. Therefore, it would appear that, for the parameters given, delocalized excitations do not have a profound effect on the timescale over which oscillatory dynamics is observed.

\subsection{Role of the localised vibrational Mode}

\begin{figure}[t!]
\subfigure[]{
\includegraphics[width=90mm]{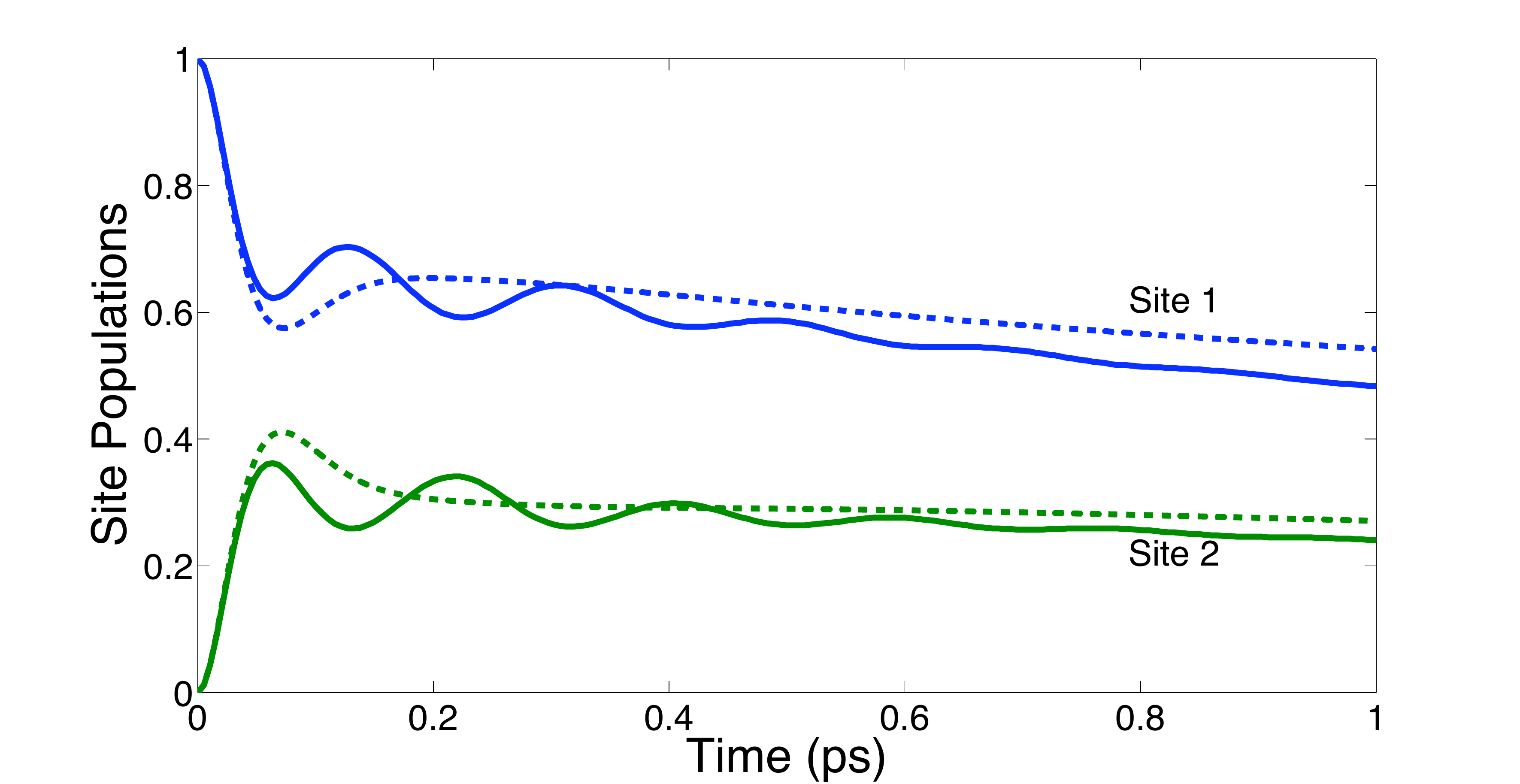}
}
\subfigure[]{
\includegraphics[width=90mm]{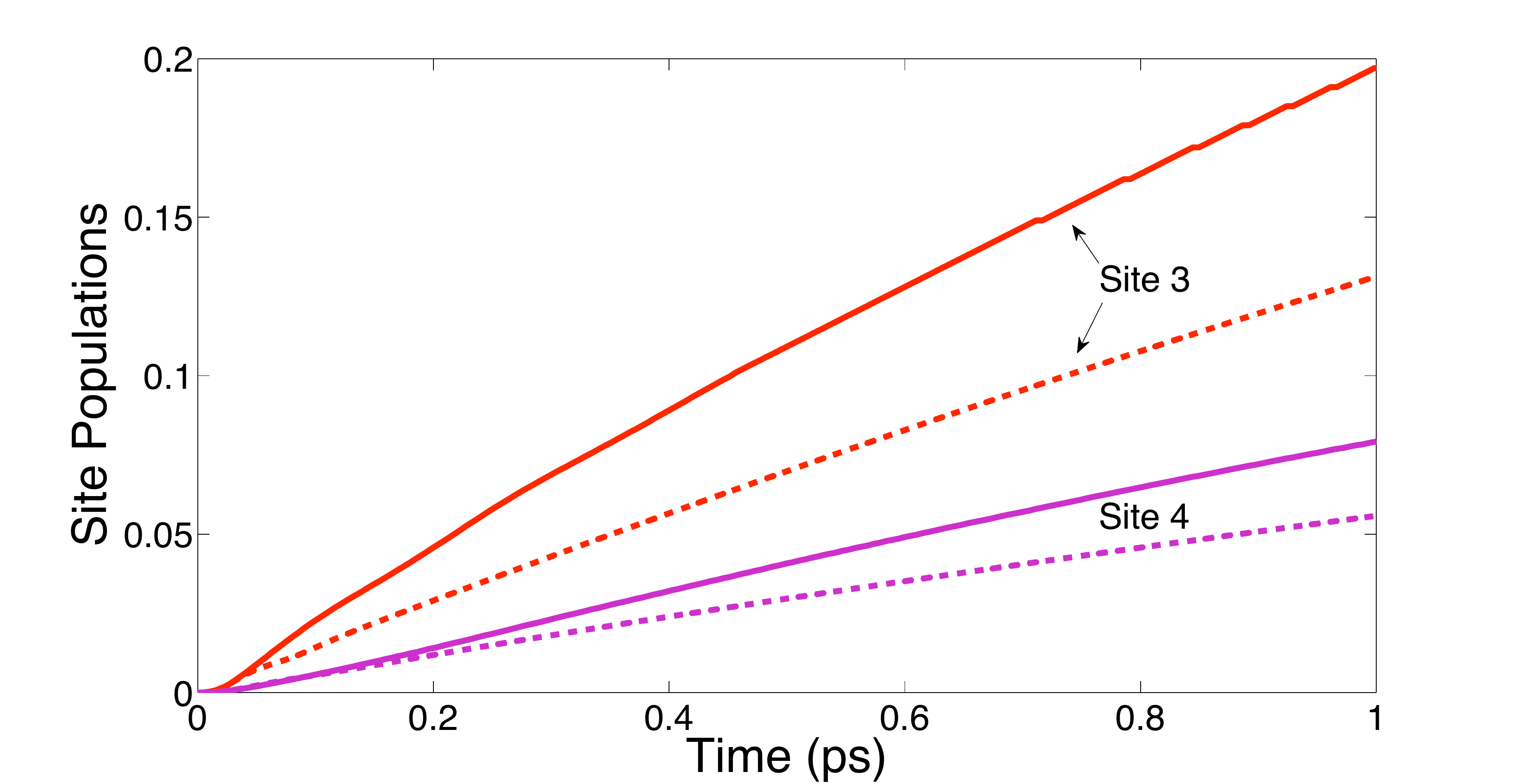}
}
\caption{Population dynamics of the FMO subsystem with (solid) and without (dashed) the localised vibrational mode. The dynamics presented assumes an initial state localised on site 1.}
\label{fig:NM_do_localised}
\end{figure}

\begin{figure}[t!]
\includegraphics[width=90mm]{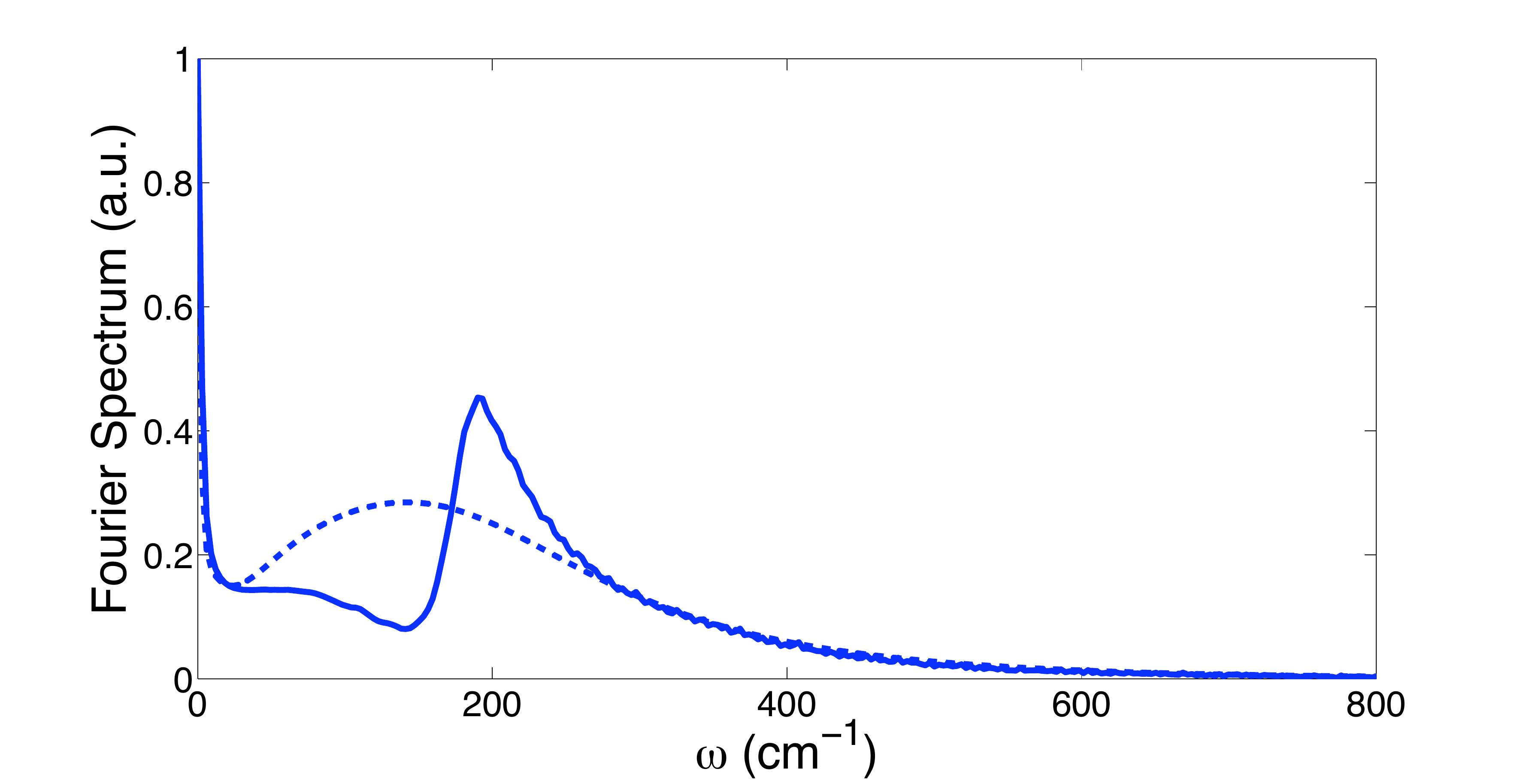}
\caption{Fourier transform spectrum of the population dynamics of site 1 in the absence (dashed) and presence (solid) of the localised vibrational mode.}
\label{fig:fouriertransform}
\end{figure}

In this section we explore the effect of the localised mode on the population dynamics and illustrate how this formalism allow us to identify the vibronic or electronic origin of the observed oscillatory dynamics.  Figure \ref{fig:NM_do_localised} presents the full non-Markovian dynamics in the absence and presence of the broadened localised mode. An important characteristic of this mode is that its energy matches the energy transitions between the highest excitonic eigenstates. This resonant condition leads to a dramatic effect on the site population dynamics as it not only enhances oscillations in the probabilities of having sites 1 and 2 excited, but also it increases the rate of energy transfer to lower energy sites 3 and 4.  We have already seen that when considering the full spectral density, including the broadened localised mode, there are strong long-lasting oscillations in the populations of sites 1 and 2. However, if the localised energy mode is neglected, such oscillations are not present. This agrees well with recent results of Prior \textit{et al.} \cite{prior10} who have predicted, using time-adaptive density matrix renormalisation methods, similar strong enhancement of coherent oscillations upon the inclusion of a localised mode. In Figure \ref{fig:fouriertransform} we present the Fourier transform of the population of site 1. In the presence of the localised mode we clearly see a strong peak at approximately $180~\textrm{cm}^{-1}$, corresponding to the energy of the localised mode. Therefore, we may associate the observed oscillations to a vibronic-induced effect. Meanwhile, in the absence of the localised mode we observe a very broad peak centred at approximately $150~\textrm{cm}^{-1}$, corresponding to the energy difference between the two highest renormalized electronic eigenstates. The broad nature of the peak is associated to very-short lived oscillatory dynamics.

\section{Markovian and secular approximations in the polaron frame }
\label{sec:4}
In general terms, for open quantum systems, non-Markovian behaviour describes a system's dynamics that is largely affected by bath memory effects and the associated system-bath correlations. Given the variety of approaches that one can follow to describe such non-Markovian phenomena, in the recent years there have been an increased number of works attempting to formally derive a measure for the degree of non-Markovianity of a quantum dynamics \cite{breuer09,wolf08,rivas10}. In particular, as it was discussed in Ref. \cite{breuer09}, while Markovian processes tend to continuously reduce the distinguishability between the dynamcal evoluiton of two different initial states, non-Markovian dynamics exhibit periods in which there is a growth of this distinguishability. This fundamental difference between Markovian and non-Markovian behaviour is illustrated in this section where we show that an interesting aspect of the many-site polaron theory is that Makovian evolutions in the polaron frame can actually capture some non-Markovian phenomena when transforming back to the lab frame.  Similarly, we will discuss that a secular approximation in the polaron frame does not necesarily imply that eigenstate populations and coherences in the lab frame are decoupled. 
\newline

\subsection{Markovian approximation} 
\label{sec:4a}

In order to consider the Markovian dynamics in the polaron frame we begin by assuming the Born approximation which implies that the system and bath states factorise at all times in the transformed frame i.e. $\tilde{\chi}(t) = \tilde{\rho}(t) \otimes \tilde{\rho}_{B}$ . Here the bath $\tilde{\rho}_{B}$ is assumed to be in thermal equilibrium in the polaron frame at all times. As a result, we see straightaway that this leads to the inhomogeneous terms given in Eq.(\ref{eq:Itotal}) to evaluate to zero. In conjuction with the Born approximation, we next assume that  the bath relaxes on a time scale much shorter than the characteristic timescale of the system's evolution i.e. Born-Markov approximation. Therefore, we may extend the upper-limit of the integrations in the rates in equation (\ref{eq:homogeneousrates}) to infinity. Now, the rates still have an explicit reference to the starting time $t=0$. This dependence on the past can be made explicit by making the substitution $s\rightarrow t-s$. The resulting Markovian master equation is $\frac{d\tilde{\rho}(t)}{dt} = \mathcal{R}_{M}(t) \tilde{\rho}(t)$ where the Markovian super-operator $\mathcal{R}_{M}(t)$ is defined as

\begin{eqnarray}\label{eq:markov}
\mathcal{R}_{M}(t)\tilde{\rho}(t)&{} ={}& - \sum_{\alpha \beta \mu \nu} \Big( \Gamma_{\alpha\beta,\mu\nu}^{(1)} e^{i (\epsilon_{\alpha\beta} + \epsilon_{\mu\nu}) t}  \big[S_{\alpha\beta}, S_{\mu\nu} \tilde{\rho}(t) \big] \nonumber\\ && ~~~~~~~~ \:{+} \Gamma_{\alpha\beta,\mu\nu}^{(2)} e^{i (\epsilon_{\beta\alpha} + \epsilon_{\mu\nu}) t} \big[S_{\alpha\beta}^{\dagger}, S_{\mu\nu} \tilde{\rho}(t) \big] \nonumber\\ && ~~~~~~~~ \:{+} \Gamma_{\alpha\beta,\mu\nu}^{(3)} e^{i (\epsilon_{\alpha\beta} + \epsilon_{\nu\mu}) t} \big[ S_{\alpha\beta}, S_{\mu\nu}^{\dagger} \tilde{\rho}(t) \big]  \nonumber\\&& ~~~~~~~~ \:{+} \Gamma_{\alpha\beta,\mu\nu}^{(4)} e^{i (\epsilon_{\beta\alpha} + \epsilon_{\nu\mu}) t} \big[ S_{\alpha\beta}^{\dagger}, S_{\mu\nu}^{\dagger} \tilde{\rho}(t) \big] \nonumber\\ 
&& ~~~~~~~~ \:{+} {\rm h.c.} \Big).
\end{eqnarray}

\noindent The time-independent, Markovian rates are:

\begin{eqnarray}
\Gamma_{\alpha\beta,\mu\nu}^{(1)} &{}={}& \int_{0}^{\infty} ds ~e^{-i\epsilon_{\mu\nu}s} ~C_{\alpha\beta,\mu\nu}^{(1)}(s), \nonumber\\
\Gamma_{\alpha\beta,\mu\nu}^{(2)} &{}={}& \int_{0}^{\infty} ds ~e^{-i\epsilon_{\mu\nu}s} ~C_{\alpha\beta,\mu\nu}^{(2)}(s), \nonumber\\
\Gamma_{\alpha\beta,\mu\nu}^{(3)} &{}={}& \int_{0}^{\infty} ds ~e^{-i\epsilon_{\nu\mu}s} ~C_{\alpha\beta,\mu\nu}^{(3)}(s), \nonumber\\
\Gamma_{\alpha\beta,\mu\nu}^{(4)} &{}={}& \int_{0}^{\infty} ds ~e^{-i\epsilon_{\nu\mu}s} ~C_{\alpha\beta,\mu\nu}^{(4)}(s),
\label{eq:markovianrates}
\end{eqnarray}

This describes the Markovian dynamics of electronic degrees of freedom {\it dressed} with a phonon reservoir. Notice, however, that separability of the system-plus-bath state in the polaron frame does not imply separability in the lab frame, as the transformation back to the lab frame converts the dressed electronic system into bare electronic degrees of freedom correlated to the vibronic  degrees of freedom through the appropriate displacements.  We therefore expect that the Markovian polaron theory should be able, under certain conditions, to capture some non-Markovian effects in the lab frame. To explore this idea, we  investigate signatures of non-Markovianity during the excitation dynamics.

A number of recent papers have proposed several measures to quantify the non-Narkovian character of quantum dynamics by quantifying the deviation of  dynamical evolution from a Markovian limit \cite{breuer09,wolf08,rivas10}. One particularly simple measure proposed by Breuer et al. \cite{breuer09} is based on the distinguishability of states as measured by the trace distance \cite{nielsen00}. The trace distance between two states $\rho_{1}$ and $\rho_{2}$ is defined as:

\begin{equation}
D(\rho_{1}(t),\rho_{2}(t)) = \frac{1}{2} \rm{tr} |\rho_{1}(t) - \rho_{2}(t)| ,
\end{equation}

\noindent where $|A| = \sqrt{A^{\dagger} A}$. It can be shown that all trace preserving completely positive maps are contractions of $D$. Therefore, all Markovian processes lead to a continuous reduction of the distinguishability between different states and the loss of distinguishability may be interpreted as the irreversible flow of information from the system to the environment \cite{breuer09}. In contrast, a non-Markovian process may be characterised by periods of increasing distinguishability during the evolution where the rate of change of $D(\rho_{1}(t),\rho_{2}(t)) $ acquires positive values.

\begin{figure}[t!]
\subfigure[Trace Distance Measure]{
\includegraphics[width=90mm]{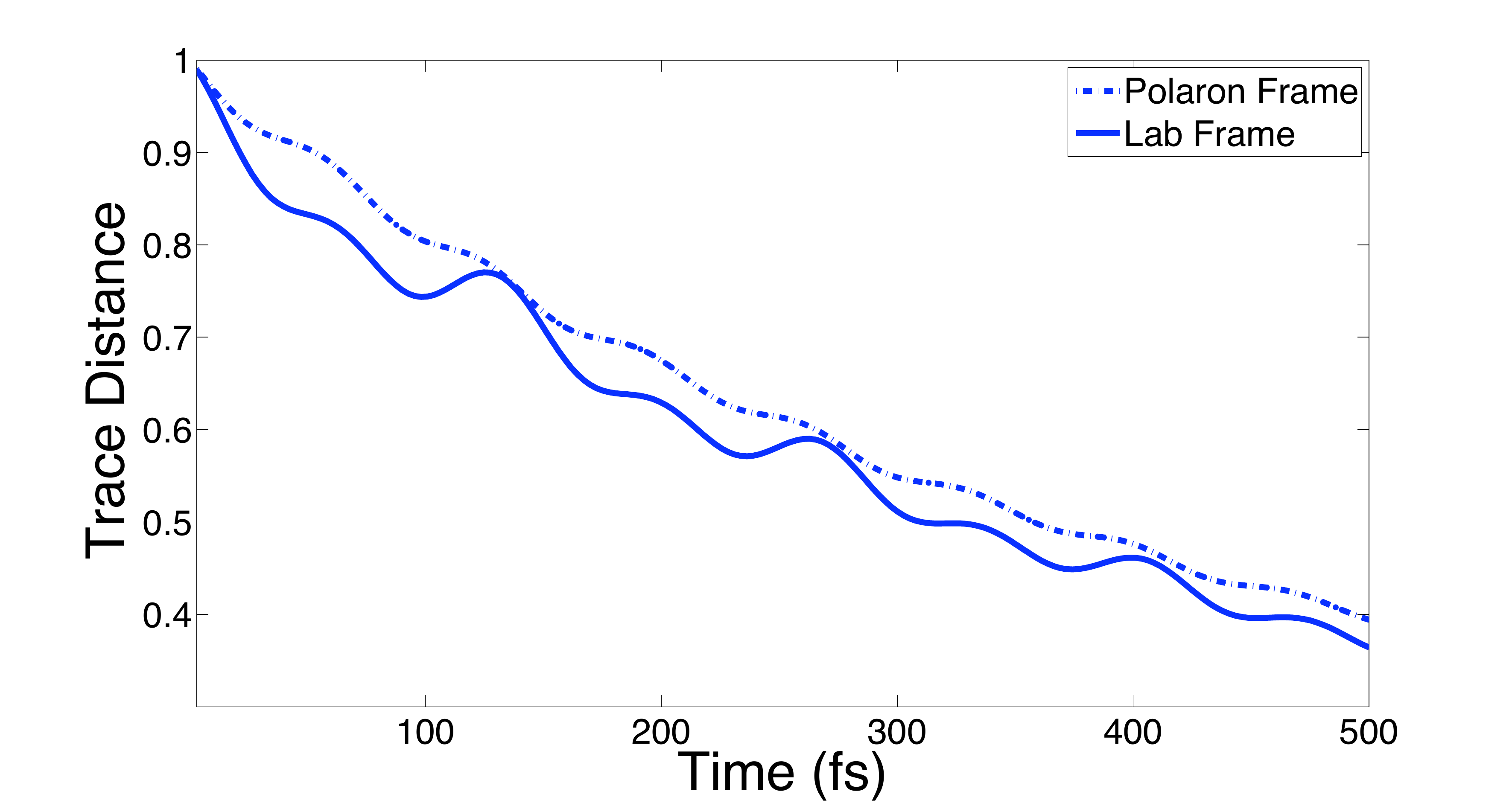}
}
\subfigure[Derivative of Trace Distance Measure]{
\includegraphics[width=90mm]{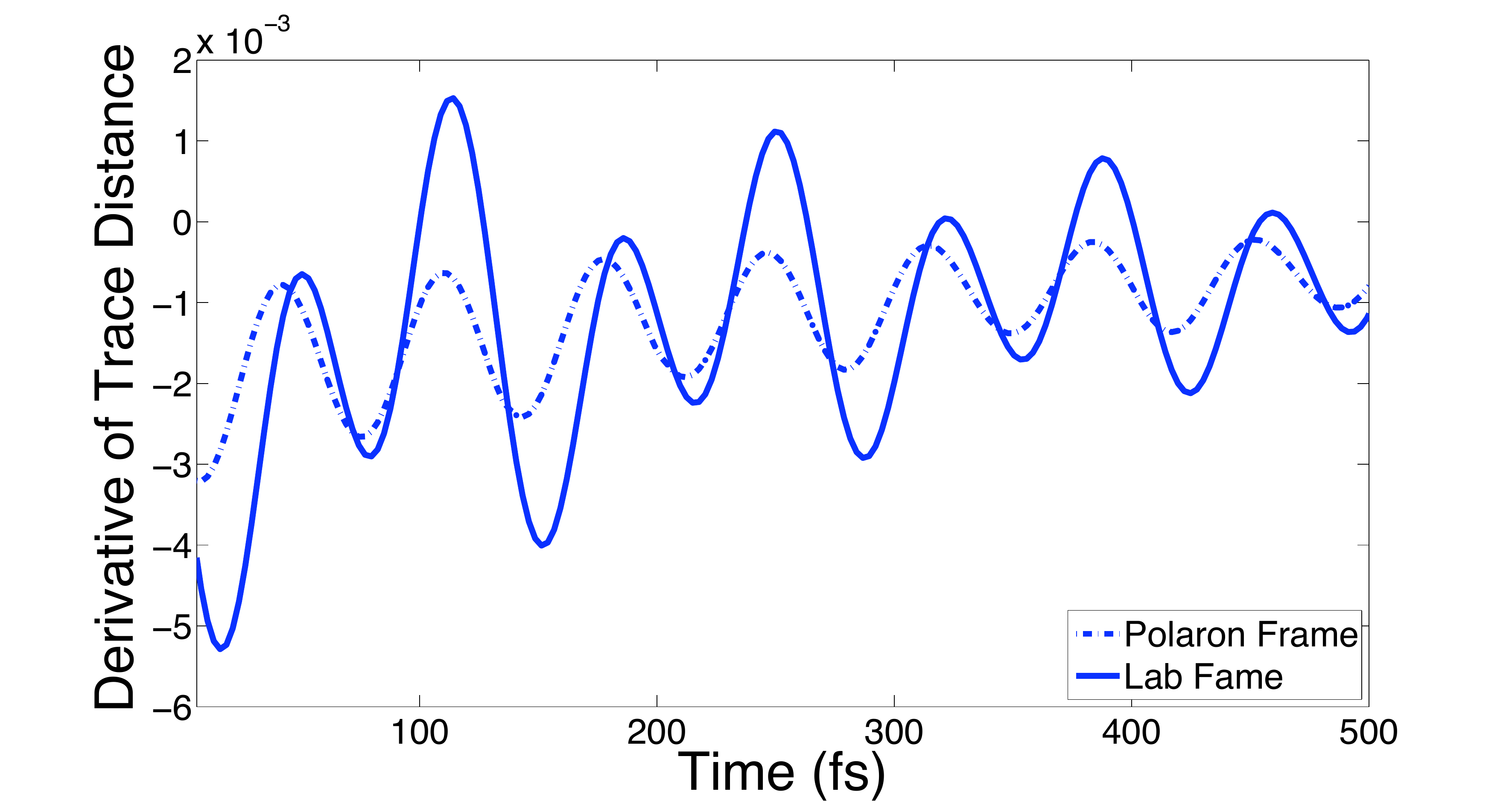}
}
\caption{Dynamics of the trace distance measure and its derivative using a Markovian polaron master equation. Presented are the measures evaluated in the polaron frame (dash-dotted) and the lab frame (solid).}
\label{fig:tracedistance}
\end{figure}

To illustrate these ideas in a specific example, we consider the same four-site subsystem of FMO. However, in order to allow us to make a valid Markovian approximation, we consider only the continuous component of the spectral density (see Eq. (\ref{eq:sd})) and ignore the localised mode that inherently leads to non-Markovian dynamics. Furthermore, we assume that the characteristic bath frequencies $\omega_{1,2}$ are increased by a factor of 10, while simultaneously the weightings $s_{1,2}$ are reduced by the same factor. This ensures that the bath correlation functions decay rapidly, while still maintaining the same reorganisation energy $\lambda=\int d\omega \, \omega^{-1} J_0(\omega)$ characterising the strength of the system-bath interaction.

Taking  $\rho_1$ and $\rho_2$ as the states where excitation is initially localised on sites 1 or 2 respectively,  figure \ref{fig:tracedistance}(a) depicts  the trace distance between $\rho_1$ and $\rho_2$ as a function of time, evaluated in both the polaron and lab frames. Although oscillations of the trace distance are observed within both frames, we notice that in the polaron frame the trace distance is always decreasing, while in the lab frame there are periods where this measure increases. To illustrate this point further, figure \ref{fig:tracedistance}(b) presents the derivative of the trace distance measure. Clearly, in the lab frame there are intervals with positive derivative denoting  an increase in distinguishability and thus suggesting a non-Markovian process. However, in the polaron frame the derivative is always strictly negative confirming the Markovianity of the dynamics within this frame. Therefore, as expected, Markovian dynamics within the polaron frame can give rise to non-Markovian evolution in the original lab frame. \newline

\subsection{Secular approximation} 
\label{sec:4b}

When the exponential terms of the form $\rm{exp} [ i (\omega + \omega') t]$ in the superoperator given in Eq. \ref{eq:markov} average to zero on the timescale relevant to bath relaxation processes, the secular approximation can be made. In that case, only the terms where  $\omega+\omega'=0$ are retained and the relaxation super-operator 
becomes
\begin{eqnarray}
\mathcal{R}_{S}(t)\tilde{\rho}(t)&{} ={}& - \sum_{\alpha \beta} \Big( \Gamma_{\alpha\beta,\beta\alpha}^{(1)} \big[S_{\alpha\beta}, S_{\beta\alpha} \tilde{\rho}(t) \big] \nonumber\\ && ~~~~~~ \:{+} \Gamma_{\alpha\beta,\alpha\beta}^{(2)} \big[S_{\alpha\beta}^{\dagger}, S_{\alpha\beta} \tilde{\rho}(t) \big] \nonumber\\ && ~~~~~~ \:{+} \Gamma_{\alpha\beta,\alpha\beta}^{(3)} \big[ S_{\alpha\beta}, S_{\alpha\beta}^{\dagger} \tilde{\rho}(t) \big]  \nonumber\\ && ~~~~~~ \:{+} \Gamma_{\alpha\beta,\beta\alpha}^{(4)} \big[ S_{\alpha\beta}^{\dagger}, S_{\beta\alpha}^{\dagger} \tilde{\rho}(t) \big] \nonumber\\ 
&& ~~~~~~ \:{+} {\rm h.c.} \Big).
\end{eqnarray}

\noindent An important consequence of the secular approximation is that in the polaron frame, eigenstate populations and coherences are decoupled from each other. To show this consider the evolution of the expectation value $\tilde{\rho}_{\mu\nu}(t)=\langle \mu | \tilde{\rho}(t) | \nu \rangle$, where $|\mu\rangle$ and $|\nu\rangle$ denote renormalized excitonic eigenstates. The population $\tilde{\rho}_{\mu\mu}(t)$ within the polaron frame evolves according to:

\begin{eqnarray}
\frac{d \tilde{\rho}_{\mu\mu}(t)}{dt} = - \sum_{\alpha} (\Gamma_{\mu\alpha,\alpha \mu}^{(1)} + \Gamma_{\alpha \mu,\alpha \mu}^{(2)} + \Gamma_{\mu\alpha,\mu\alpha}^{(3)} + \Gamma_{\alpha \mu,\mu \alpha }^{(4)} ) \tilde{\rho}_{\mu\mu}(t) ~~ \nonumber\\
 + \sum_{\alpha} (\Gamma_{\alpha \mu,\mu \alpha }^{(1)} +  \Gamma_{\mu\alpha,\mu\alpha}^{(2)} + \Gamma_{\alpha \mu,\alpha \mu}^{(3)} +  \Gamma_{\mu\alpha,\alpha \mu}^{(4)} ) \tilde{\rho}_{\alpha\alpha}(t) \nonumber\\
 + ~\textrm{h.c.} ~~~~~~~~~~~~~~~~~~~~~~~~~~~~~~~~~~~~~~~~~~~~~~~~~~~~~~~ \nonumber\\
\label{eq:secularpops} 
\end{eqnarray}

\noindent Meanwhile, the off-diagonal matrix element $\tilde{\rho}_{\mu\nu}(t)$ evolve as follows:

\begin{eqnarray}
\frac{d\tilde{\rho}_{\mu\nu}(t)}{dt}= - \sum_{\alpha} (\Gamma_{\mu\alpha,\alpha \mu}^{(1)} + \Gamma_{\alpha \mu,\alpha \mu}^{(2)} + \Gamma_{\mu\alpha, \mu \alpha}^{(3)} + \Gamma_{\alpha \mu,\mu \alpha}^{(4)}) \tilde{\rho}_{\mu\nu}(t)  \nonumber\\
+ \textrm{h.c.} ~~~~~~~~~~~~~~~~~~~~~~~~~~~~~~~~~~~~~~~~~~~~~~~~~
\end{eqnarray}

\begin{figure}[t!]
\subfigure[Population of the two highest-energy renormalized excitonic states $E_1$ and $E_2$.]{
\includegraphics[width=90mm]{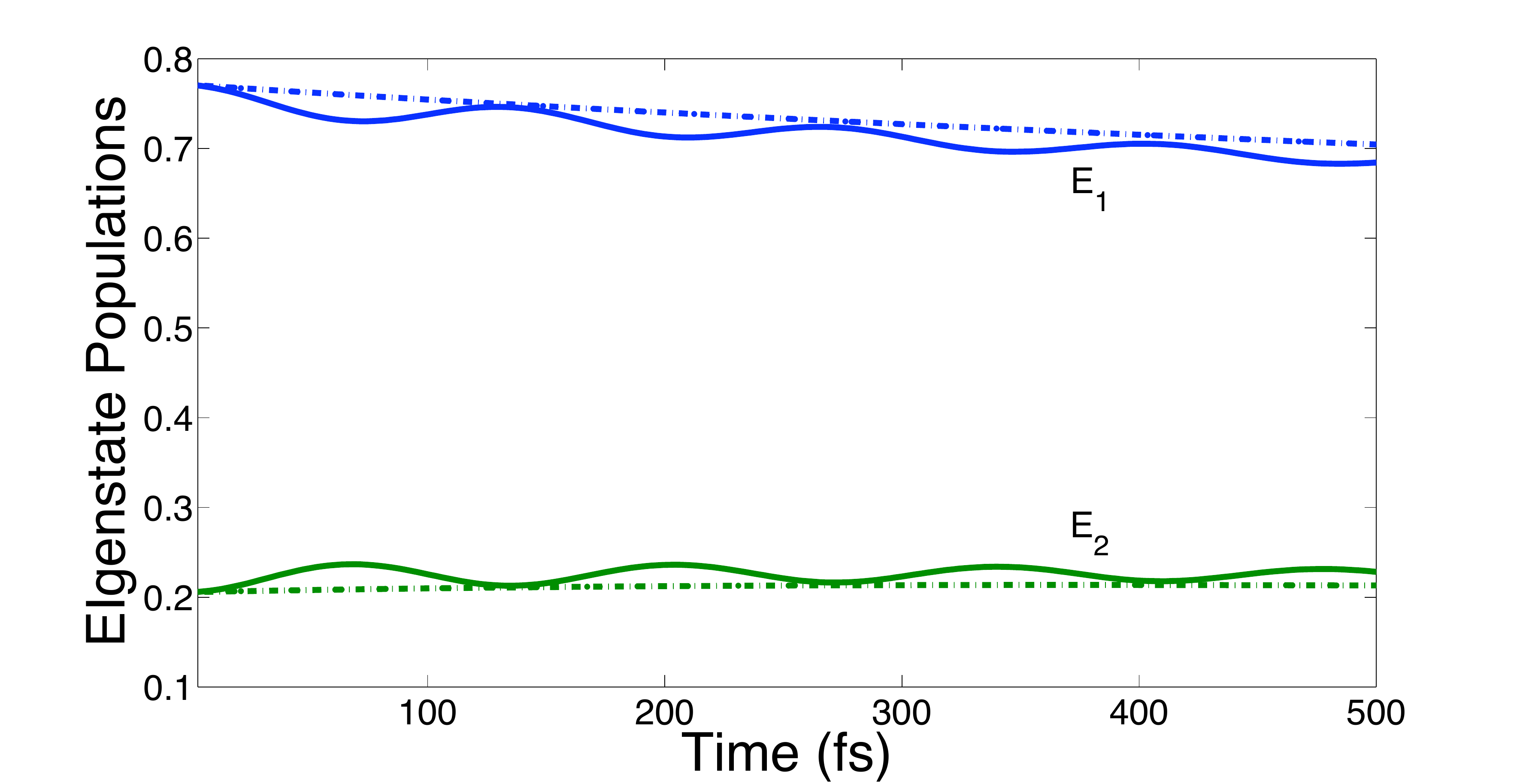}
}
\subfigure[Population of the two lowlying renormalized excitonic states $E_3$ and $E_4$]{
\includegraphics[width=90mm]{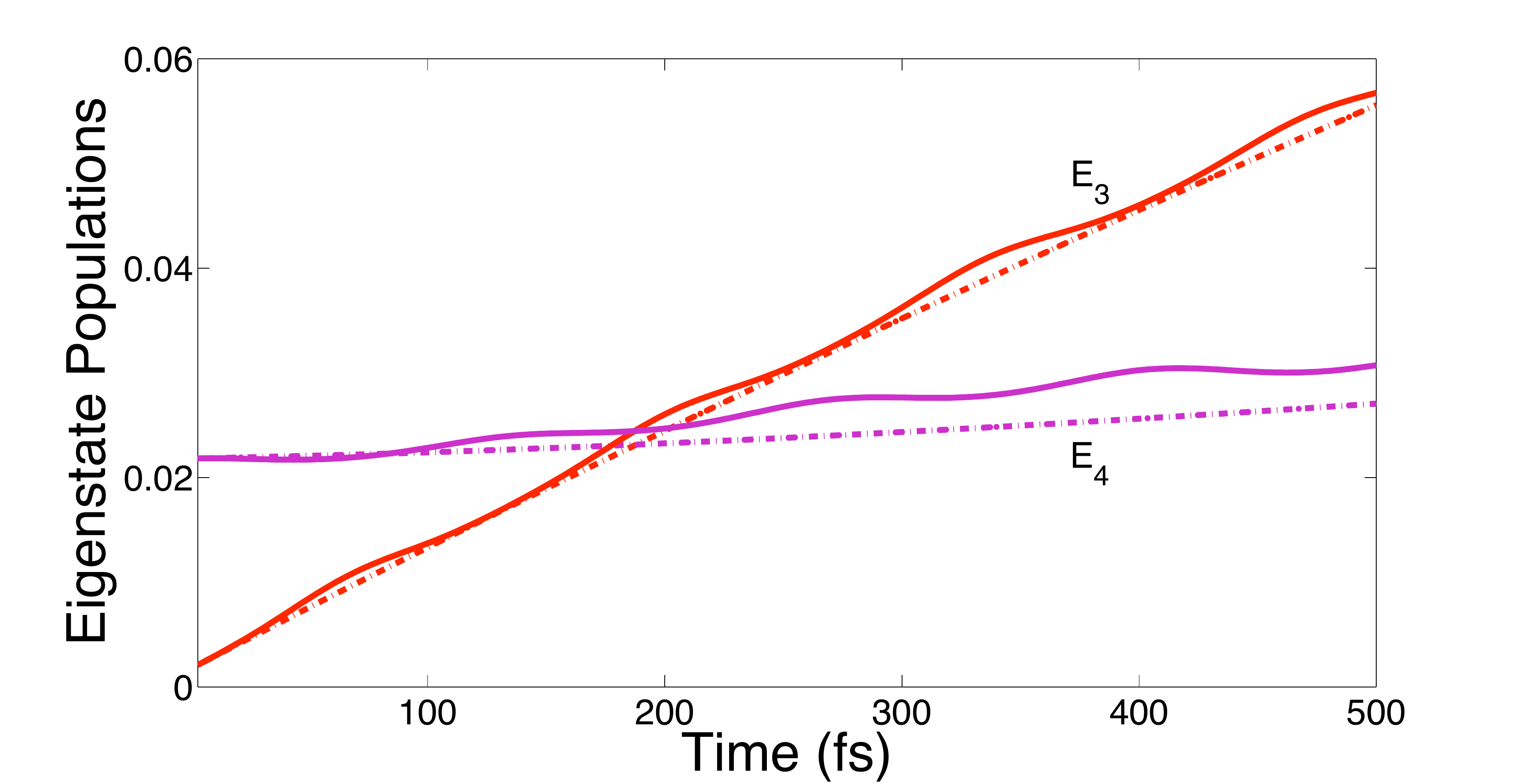}
}
\caption{Dynamics of the eigenstate populations using the secular Markovian polaron master equation evaluated in the polaron frame (dash-dotted) and the lab frame (solid).}
\label{fig:secular}
\end{figure}

Notice that eigenstate populations obey a simple Pauli master equation. Hence, no oscillatory dynamics is expected for renormalized exciton populations in the polaron frame. However, since the polaron transformation does not commute with the rotation between the site basis and the renormalised eigenstate basis, there is a mixing of populations and coherences upon transforming back into the lab frame, and hence beating of the population of renormalised excitonic states in the lab frame should be observed. To illustrate this effect, we consider the same modified FMO system as described in Sec. \ref{sec:4a}. Figure \ref{fig:secular}  shows the dynamics of the  renormalized exciton eigenstate populations in both polaron and lab frames, as predicted by a polaron master equation with secular approximation for an excitation initially localised on site 1. While population transfer proceeds in an inchoherent manner in the polaron frame, in the original lab frame we observe a clear oscillatory behaviour of the populations of all four renormalized excitonic eigenstates, confirming that populations and coherences are coupled. 
\newline

\noindent\textbf{Weak-Coupling Limit:} In the limit that the system-bath coupling is sufficiently weak we may approximate the Markovian rates given in equation (\ref{eq:markovianrates}) by expanding the correlation functions $C_{\alpha\beta,\mu\nu}^{(i)}(t)$ in powers of $\mathcal{K}_{mn,pq}(t)$. To first order, we find

\begin{eqnarray}
C_{\alpha\beta,\mu\nu}^{(1)}(t) = C_{\alpha\beta,\mu\nu}^{(4)}(t) &\approx& -C_{\alpha\beta,\mu\nu}^{(W)}(t), \nonumber\\
C_{\alpha\beta,\mu\nu}^{(2)}(t) = C_{\alpha\beta,\mu\nu}^{(3)}(t) &\approx& C_{\alpha\beta,\mu\nu}^{(W)}(t),
\end{eqnarray}

\noindent where

\begin{equation}\label{CWeak}
C_{\alpha\beta,\mu\nu}^{(W)}(t)=\sum_{\langle mn \rangle} \sum_{\langle pq \rangle} \mathcal{U}_{\alpha\beta\mu\nu}^{mnpq}\tilde{\beta}_{mn} \tilde{\beta}_{pq} \mathcal{K}_{mn,pq}(t),
\end{equation}
is the weak-coupling bath correlation function.

The rates in Eq.~(\ref{eq:markovianrates}) can now be evaluated in a straightforward manner by substituting in the form of the $\mathcal{K}_{mn,pq}(t)$ (as defined in Appendix \ref{sec:a1}) into Eq.~(\ref{CWeak}), and making use of the relation $\int_0^{\infty}dt e^{i\omega t}=\pi\delta(\omega)+iP(1/\omega)$, where $P$ denotes the principal value, to perform the integrals over time. Hence, in the weak coupling limit, all four rates can be written in a generic form
\begin{eqnarray}
\Gamma_{\alpha\beta,\mu\nu}^{(W)}(\epsilon)&{}={}& \int_{0}^{\infty} ds~e^{-i\epsilon s} ~C_{\alpha\beta,\mu\nu}^{(W)}(s), \nonumber\\
&{}={}&\gamma_{\alpha\beta,\mu\nu}(\epsilon)-iS_{\alpha\beta,\mu\nu}(\epsilon).
\end{eqnarray}
Here
\begin{eqnarray}
\gamma_{\alpha\beta,\mu\nu}(\epsilon)&{}={}&\frac{\pi}{2}\sum_{\langle mn \rangle} \sum_{\langle pq \rangle} \mathcal{U}_{\alpha\beta\mu\nu}^{mnpq}\tilde{\beta}_{mn} \tilde{\beta}_{pq}\frac{J(\epsilon)}{\epsilon^2}\lambda_{mn,pq}(\epsilon)\nonumber\\
&&\:{\times}\left(\coth{(\beta\epsilon/2)}-1\right)
\end{eqnarray}
is the expected single-phonon relaxation rate, while
\begin{eqnarray}
S_{\alpha\beta,\mu\nu}(\epsilon)&{}={}&\sum_{\langle mn \rangle} \sum_{\langle pq \rangle} \mathcal{U}_{\alpha\beta\mu\nu}^{mnpq}\tilde{\beta}_{mn} \tilde{\beta}_{pq}\nonumber\\
&&\:{\times}P\int_0^{\infty}d\omega\frac{J(\omega)}{\omega^2}\frac{\lambda_{mn,pq}(\omega)}{\omega^2-\epsilon^2}(\omega-\epsilon\coth{\beta\omega/2})\nonumber\\
\end{eqnarray}
is the associated bath-induced energy shift. Hence, provided that we may legitimately perform the expansion in $\mathcal{K}_{mn,pq}(t)$, we see that our master equation should correctly capture the expected Redfield dynamics of the system in the weak-coupling limit.
\newline

\noindent\textbf{F\"orster Limit:} In the opposite regime of strong system-bath interaction, the bath renormalisation factors $\beta_{mn}$ tend to zero. In this limit, electronic couplings are renormalised to zero and the eigenstate basis is simply the site basis ($u_{m\alpha}=\delta_{m\alpha}$).  It can be shown that when $\beta_{mn} \rightarrow 0$, only the rates $\Gamma_{\alpha\beta,\alpha\beta}^{(2)}$ and $\Gamma_{\alpha\beta,\alpha\beta}^{(3)}$ have non-zero contribution and evaluate to:
\begin{eqnarray}
\Gamma_{\alpha\beta,\alpha\beta}^{(2)} &=& \sum_{\langle mn\rangle} V_{mn}^{2} \delta_{m\alpha} \delta_{n \beta} \Gamma^{(S)}(\epsilon_{\alpha \beta}) \nonumber\\
\Gamma_{\alpha\beta,\alpha\beta}^{(3)} &=& \sum_{\langle mn\rangle} V_{mn}^{2} \delta_{m\alpha} \delta_{n\beta} \Gamma^{(S)}(-\epsilon_{\alpha\beta}),
\end{eqnarray}

\noindent where we have defined the strong coupling limit rate as

\begin{equation}
\Gamma^{(S)}(\omega) = \int_{0}^{\infty} ds e^{-i\omega s} e^{-\mathcal{K}_{mn,mn}(0) +  \mathcal{K}_{mn,mn}(s)}.
\end{equation}

\noindent It can be shown that in this limit, the site populations follow the Pauli master equation presented in Eq. (\ref{eq:secularpops}). On substituting in the forms of the rates $\Gamma_{\alpha\beta,\alpha\beta}^{(2)}$ and $\Gamma_{\alpha\beta,\alpha\beta}^{(3)}$ from above, and after performing some simple manipulations, we arrive at the strong system-bath coupling master equation:

\begin{eqnarray}
\frac{d \tilde{\rho}_{nn}(t)}{dt} =  \sum_{m\neq n} V_{nm}^{2} \mathit{Re}[\Gamma^{(S)}(\epsilon_{nm})] \tilde{\rho}_{mm}(t) \nonumber\\ - \sum_{m\neq n} V_{mn}^{2} \mathit{Re}[\Gamma^{(S)}(\epsilon_{mn})]  \tilde{\rho}_{nn}(t).
\end{eqnarray} 
Notice that the above is exactly  the incoherent F\"orster dynamics with rates $\mathit{V_{mn}^{2}Re}[ \Gamma^{(S)}(\epsilon_{mn})]$ corresponding to the F\"orster transfer rate from site $m$ to site $n$.

\section{Concluding remarks}
\label{sec:5}

It is currently  of much interest to develop theories of multichromopore electronic excitation dynamics capable of bridging the gap between the limiting cases of weak and strong exciton-phonon coupling,  whilst  remaining computationally tractable as the number of chromophores increases.  In this context, modified perturbative methodologies, as the one presented here, provide a valuable alternative to exact treatments.  In particular, this paper generalizes the polaron-modified perturbative master equation originally presented for a donor-aceptor pair to the case of multichromophore excitation dynamics. Explicit expressions for the homogeneous and inhomogeneous super-operators for an arbitrary non-equilibrium bath initial state have been presented. To illustrate the scope of this many-site theory, we have investigated electronic excitation dynamics in a four-site subsystem of the FMO complex under the influence of a structured phonon bath that includes a localised vibrational mode. Our results indicate that in this example the non-equilibrium bath dynamics, captured by the inhomogeneous contribution, is crucial to give an accurate account of the origin and time scale of oscillations on the ultrafast scale. In particular, we show how the theory can describe the enhancement and modification of the oscillatory dynamics due to strong coupling to a localised, yet broadened vibrational mode. In a separate publication we will be describing in detail how the interplay between electronic coherence and localised vibrational modes can generate rich behaviour of transfer of excitions in light harvesting systems. The ability to understand this interplay will give insights into the role of coherent dynamics in systems where the excitonic transitions may be resonant with localised vibrational modes.

In addition to calculating site population dynamics in the lab frame, we have outlined a framework for evaluating all possible electronic observables in the lab frame. This will allow a full reconstruction of the lab frame density matrix for excitation dynamics and hence enable comparisons with experimental observations. However, explicit calculation of non-equilibrium contributions to the expected values of certain electronic operators are beyond the scope of this paper, so we have presented here a zeroth-order approximation of such contributions and leave the full calculation for a forthcoming publication. 

Finally, we have presented the Markovian and secular approximations of the multichromophore polaron-modified master equation. Despite these approximations being satisfied in the polaron frame, we have shown that both approximations do not necessarily have to hold in  the untransformed lab frame. This suggests that there is scope for Markov polaron master equations to capture aspects of non-Markovian or non-secular dynamics in the lab frame in particular parameter regimes.\\

Note: on completation of this work we were made aware of a similar work by Jang\cite{jang11}. Our approaches are similar in scope and objective but our analysis present complementary insights. 

\noindent \textbf{Acknowledgements}

We would like to thank Michael Thorwart for discussions and comments to our paper and to Seogjoo Jang for providing a pre-print of his manuscript while ours was under revision. AK and AOC acknowledge funding from the EPSRC (grant No. EP/G005222/1 ). AN is grateful with the EPSRC and Imperial College London for support.

\appendix

\section{Homogeneous correlations functions}
\label{sec:a1}
In order to calculate the bath correlation functions, the polaron frame bath operators are written in terms of displacement operators:

\begin{equation}
B_{mn}(t) = \prod_{\mathbf{k}} D(\delta\alpha_{\mathbf{k},mn}(t))
\end{equation}

\noindent where the displacement operator is in general defined as $D(\alpha_{\mathbf{k}})=e^{\alpha_{\mathbf{k}} b_{\mathbf{k}}^{\dagger} - \alpha_{\mathbf{k}}^{*} b_{\mathbf{k}}}$. Using the following properties of displacement operators

\begin{eqnarray}
D(\alpha_{\mathbf{k}})D(\beta_{\mathbf{k}}) = e^{(\alpha_{\mathbf{k}}\beta_{\mathbf{k}}^{*}-\alpha_{\mathbf{k}}^{*}\beta_{\mathbf{k}})/2} D(\alpha_{\mathbf{k}}+\beta_{\mathbf{k}}) \nonumber\\
\langle D(\alpha_{\mathbf{k}}) \rangle = \textrm{exp}\Big(-\frac{1}{2} |\alpha_{\mathbf{k}}|^{2} \coth(\beta\omega_{\mathbf{k}}/2)\Big) ~~~
\end{eqnarray}
the final expressions for the homogeneous bath correlation functions become

\begin{eqnarray}
\left. \begin{array}{c} 
\langle \tilde{B}_{mn}(t) \tilde{B}_{pq}(s)\rangle \\  
\langle \tilde{B}_{mn}^{\dagger}(t) \tilde{B}_{pq}^{\dagger}(s)\rangle 
\end{array} \right\} &=& \beta_{mn} \beta_{pq} (e^{-\mathcal{K}_{mn,pq}(t-s)} - 1), \nonumber\\
\nonumber\\
\left. \begin{array}{c} 
\langle \tilde{B}^{\dagger}_{mn}(t) \tilde{B}_{pq}(s)\rangle \\  
\langle \tilde{B}_{mn}(t) \tilde{B}_{pq}^{\dagger}(s)\rangle 
\end{array} \right\} &=& \beta_{mn} \beta_{pq} (e^{\mathcal{K}_{mn,pq}(t-s)} - 1), \nonumber\\
\end{eqnarray}

\noindent where the correlation function $\mathcal{K}_{mn,pq}(t)$ is defined as

\begin{eqnarray}
\mathcal{K}_{mn,pq}(t) = \int_{0}^{\infty} d\omega \frac{J(\omega)}{\omega^{2}} \lambda_{mn,pq} \big( \coth(\beta \omega/2) \cos (\omega(t)) \nonumber\\  - i \sin (\omega(t) \big), ~~~ \nonumber\\
\end{eqnarray} 

\noindent and the spatial correlation function is defined as $\lambda_{mn,pq} = \Delta_{m,p} - \Delta_{m,q} - \Delta_{n,p} + \Delta_{n,q}$. Here $\Delta_{m,p}$ describes the degree of spatial correlation between sites $m$ and $p$. For the propagating modes model of spatial correlations $\Delta_{m,p} = \Delta_{m,p}(\omega) = \textrm{sinc}(\omega d_{mn}/v_{ph}) $.

\section{Inhomogeneous correlation functions}
\label{sec:a2}
Let us consider a general initial state within the {\it lab} frame (i.e. prior to polaron tranformation): $\chi(0) = \sum_{ij} \rho_{ij}(0) \sigma_{i}^{+} \sigma_{j}^{-} \otimes \rho_{B}$, where $\rho_{B}$ denotes the thermal equilibrium bath state in the lab frame. Transforming into the polaron frame we find the initial state
\begin{equation}
\tilde{\chi}(0) = \sum_{ij} \tilde{\rho}_{ij}(0) \sigma_{i}^{+} \sigma_{j}^{-} \prod_{\mathbf{k}} \beta_{ij}^{-1} D(\alpha_{\mathbf{k},i}) \tilde{\rho}_{B} D(-\alpha_{\mathbf{k},j}).
\end{equation}
Here $\tilde{\rho}_{ij}(0) = \beta_{ij} \rho_{ij}(0)$ and denotes the $ij$'th element of the initial system density operator in the polaron frame. 

The irrelevant part of the total system-bath density matrix at time zero is then given by
\begin{eqnarray}
\mathcal{Q}\tilde{\chi}(0) &{}={}& \sum_{ij} \tilde{\rho}_{ij}(0) \sigma_{i}^{+} \sigma_{j}^{-}\prod_{\mathbf{k}} \Big( \beta_{ij}^{-1} D(\alpha_{\mathbf{k},i}) \tilde{\rho}_{B} D(-\alpha_{\mathbf{k},j}) - \tilde{\rho}_{B} \Big) \nonumber\\
&=& \sum_{ij} \tilde{\rho}_{ij}(0) \sigma_{i}^{+} \sigma_{j}^{-}Q_{ij} \tilde{\rho}_{B}.
\end{eqnarray}
Notice that we have defined $Q_{ij}\tilde{\rho}_{B}$ as the state accounting for the difference between the displaced bath and the bath thermal equilibrium in the polaron frame. 

Expectation values with respect to the state $Q_{ij}\tilde{\rho}_{B}$ can be expressed in terms of  expectation values taken with respect to the thermal equilibrium state in the polaron frame, as follows:

\begin{equation}
\langle X \rangle_{Q_{ij}\tilde{\rho}_{B}}  =  \prod_{\mathbf{k}} \langle D(-\alpha_{\mathbf{k},j}) X D(\alpha_{\mathbf{k},i}) \rangle_{\tilde{\rho}_{B}} - \langle X \rangle_{\tilde{\rho}_{B}}
\end{equation}

Using this identity and the previous properties of displacement operators, we can now calculate the various inhomogeneous correlation functions. The full expression for the correlation function appearing in the first order term of the inhomogeneous super-operator can be evaluated as:

\begin{equation}
\langle \tilde{B}_{mn}(t) \rangle_{Q_{ij}\tilde{\rho}_{B}}  = \beta_{mn} (f_{ij,mn}(t) -1).
\end{equation}

\noindent Here the correlation function $f_{ij,mn}(t)$ is defined as

\begin{eqnarray}
f_{ij,mn}(t) = e^{- \int_{0}^{\infty} d\omega \frac{J(\omega)}{\omega^{2}} \lambda_{ij,mn}\coth(\beta\omega/2)\cos(\omega t)} \nonumber\\
\times e^{ i \int_{0}^{\infty} d\omega \frac{J(\omega)}{\omega^{2}} \lambda_{ij,mn}^{'} \sin(\omega t)}~~~~~~~~
\end{eqnarray}

\noindent The spatial correlation factor $\lambda_{ij,mn}$ is as defined in Appendix \ref{sec:a1}, while a second spatial correlation function is introduced: $\lambda_{ij,mn}' =\Delta_{i,m} - \Delta_{i,n} + \Delta_{j,m} - \Delta_{j,n}$.

\noindent The correlation functions appearing in the second order inhomogeneous term are given by:

\begin{eqnarray}
\langle \tilde{B}_{mn}(t) \tilde{B}_{pq}(s) \rangle_{Q_{ij}\tilde{\rho}_{B}} ~~~~~~~~~~~~~~~~~~~~~~~~~~~~~~~~~~~~~~~~~~~~~~~\nonumber\\
= \beta_{mn}\beta_{pq} \Big( \big( f_{ij,mn}(t) f_{ij,pq}(s) - 1 \big) e^{-\mathcal{K}_{mn,pq}(t-s)} ~~~~~~ \nonumber\\ 
- f_{ij,mn}(t) - f_{ij,pq}(s) + 2 \Big), ~~~~~~~~~ \nonumber\\
\nonumber\\
\langle \tilde{B}_{mn}^{\dagger}(t) \tilde{B}_{pq}(s)  \rangle_{Q_{ij}\tilde{\rho}_{B}} ~~~~~~~~~~~~~~~~~~~~~~~~~~~~~~~~~~~~~~~~~~~~~~~\nonumber\\
= \beta_{mn}\beta_{pq} \Big( \big( f_{ij,mn}'(t) f_{ij,pq}(s) - 1 \big) e^{\mathcal{K}_{mn,pq}(t-s)} ~~~~~~ \nonumber\\ 
- f_{ij,mn}'(t) - f_{ij,pq}(s) + 2 \Big), ~~~~~~~~~ \nonumber\\
\nonumber\\
\langle \tilde{B}_{mn}(t) \tilde{B}_{pq}^{\dagger}(s) \rangle_{Q_{ij}\tilde{\rho}_{B}} ~~~~~~~~~~~~~~~~~~~~~~~~~~~~~~~~~~~~~~~~~~~~~~~\nonumber\\ 
= \beta_{mn}\beta_{pq} \Big( \big( f_{ij,mn}(t) f_{ij,pq}'(s) - 1 \big) e^{\mathcal{K}_{mn,pq}(t-s)} ~~~~~~ \nonumber\\ 
- f_{ij,mn}(t) - f_{ij,pq}'(s) + 2 \Big), ~~~~~~~~~ \nonumber\\
\nonumber\\
\langle \tilde{B}_{mn}^{\dagger}(t) \tilde{B}_{pq}^{\dagger}(s) \rangle_{Q_{ij}\tilde{\rho}_{B}} ~~~~~~~~~~~~~~~~~~~~~~~~~~~~~~~~~~~~~~~~~~~~~~~\nonumber\\ 
= \beta_{mn}\beta_{pq} \Big( \big( f_{ij,mn}'(t) f_{ij,pq}'(s) - 1 \big) e^{-\mathcal{K}_{mn,pq}(t-s)} ~~~~~~ \nonumber\\ 
- f_{ij,mn}'(t) - f_{ij,pq}'(s) + 2 \Big). ~~~~~~~~~ \nonumber\\
\end{eqnarray}

\noindent Here we have introduced a final correlation function:

\begin{eqnarray}
f_{ij,mn}'(t) &=& \big( f_{ij,mn}(t) \big)^{-1} \nonumber\\
&=& e^{\int_{0}^{\infty} d\omega \frac{J(\omega)}{\omega^{2}} \lambda_{ij,mn}\coth(\beta\omega/2)\cos(\omega t)} \nonumber\\
&& ~~~ \times e^{- i \int_{0}^{\infty} d\omega \frac{J(\omega)}{\omega^{2}} \lambda_{ij,mn}^{'} \sin(\omega t)}
\end{eqnarray}

\section{Numerical Integration}
	
For convenience we numerically solve the dynamics in the polaron frame within the eigenstate basis of the renormalised Hamiltonian in Equation \ref{eq:h0}. In this basis, we may write the polaron master equation as:

\begin{equation}
\frac{d \tilde{\rho}_{\alpha\beta}(t)}{dt} = \sum_{\mu\nu} R_{\alpha\beta,\mu\nu}(t) \tilde{\rho}_{\mu\nu}(t) + I_{\alpha\beta}(t)
\end{equation}

\noindent Here, $R_{\alpha\beta,\mu\nu}(t)$ and $I_{\alpha\beta}(t)$ are time-dependent tensors corresponding to the homogeneous and inhomogeneous superoperators respectvely. To simplify the numerics, we flatten the system density matrix to form a vector describing the state: $\uline{\rho} = (\tilde{\rho}_{11}, \tilde{\rho}_{12}, \tilde{\rho}_{13}, \dots, \tilde{\rho}_{NN})^{T}$. In this new representation, we may write the master equation in terms of the following matrix equation:

\begin{equation}
\frac{d}{dt} \uline{\rho}(t) = \uuline{R}(t) . \uline{\rho}(t) + \uline{I}(t)
\end{equation}

\noindent This matrix equation is numerically integrated using the fourth order Runge-Kutta method. 

At each time step during the numerical integration, the elements of the homogeneous matrix $\uuline{R}(t)$ and the inhomogeneous vector $\uline{I}(t)$ are determined from the expressions in Equations \ref{eq:R}, \ref{eq:I1} and \ref{eq:I2}. The most computationally intensive step in evaluating these two terms occurs in performing the integrations to calculate the time-dependent homogeneous and inhomogeneous rates. To reduce operation time, at each time step we calculate all rates first, before building up the homogeneous matrix and inhomogeneous vector. Furthermore, we notice that at each time step all the rates can all be calculated independently. Therefore, we may also utilise parallelisation algorithms to further enhance the performance of the numerical integration.

\end{document}